\documentclass[aps,pra,amsmath,amssymb,superscriptaddress]{revtex4}
\usepackage{times}
\usepackage{latexsym}
\usepackage{graphicx}
\usepackage{amsmath,amssymb}
\usepackage{tensor}
\usepackage{verbatim,bbm}
\usepackage{epstopdf}
\usepackage{xcolor}
\usepackage{color}
\usepackage[normalem]{ulem}

\DeclareMathOperator{\Tr}{Tr}

\begin{document}

\title{Characterising port-based teleportation as universal simulator of qubit channels}
\author{Jason Pereira}
\affiliation{Department of Computer Science, University of York, York YO10 5GH, UK}
\author{Leonardo Banchi}
\affiliation{Department of Physics and Astronomy, University of Florence, via G. Sansone 1, I-50019 Sesto Fiorentino (FI), Italy}
\affiliation{INFN Sezione di Firenze, via G. Sansone 1, I-50019, Sesto Fiorentino (FI), Italy.}
\author{Stefano Pirandola}
\affiliation{Department of Computer Science, University of York, York YO10 5GH, UK}

\begin{abstract}
Port-based teleportation (PBT) is a teleportation protocol that employs a number of Bell pairs and a joint measurement to enact an approximate input-output identity channel. Replacing the Bell pairs with a different multi-qubit resource state changes the enacted channel and allows the PBT protocol to simulate qubit channels beyond the identity. The channel resulting from PBT using a general resource state is consequently of interest. In this work, we fully characterise the Choi matrix of the qubit channel simulated by the PBT protocol in terms of its resource state. We also characterise the PBT protocol itself, by finding a description of the map from the resource state to the Choi matrix of the channel that is simulated by using that resource state. Finally, we exploit our expressions to show improved simulations of the amplitude damping channel by means of PBT with a finite number of ports.
\end{abstract}

\maketitle

\section{Introduction}
Quantum teleportation \cite{bennett_teleporting_1993,braunstein_teleportation_1998,pirandola_advances_2015} is a powerful tool in quantum information \cite{nielsen_quantum_2011,watrous_theory_2018,bengtsson_geometry_2006,holevo_quantum_2019,braunstein_quantum_2005,weedbrook_gaussian_2012,andersen_hybrid_2015}. Teleportation protocols utilise entanglement between quantum states held by a sender and a receiver to transmit a state. The resulting quantum channel mapping the sent state to the received state is determined by the protocol used and by the resource state held by the sender and receiver prior to the protocol being enacted. Such protocols have applications in quantum communications protocols (for example, superdense coding \cite{nielsen_quantum_2011}) as well as in quantum computing (quantum gate teleportation \cite{gottesman_demonstrating_1999}), and they can be used as a mathematical tool for the simulation of quantum channels \cite{pirandola_fundamental_2017,pirandola_theory_2018} and quantum networks~\cite{pirandola_end--end_2019,pirandola_bounds_2019}.

The standard teleportation protocol, as proposed by Bennett et al.~\cite{bennett_teleporting_1993}, uses a shared (between the sender and the receiver) two-qubit state. A measurement is performed on the sender's qubit and the qubit to be teleported, projecting the pair of qubits onto a Bell state. Based on the result of this measurement, one of the four Pauli operators (including the identity) is applied to the receiver's state. The quantum channel resulting from teleportation using this protocol depends on the resource state used. This protocol has limitations, however, as it is only able to simulate Pauli channels~\cite{bowen_teleportation_2001}. This stems from the fact that the Pauli operators, which are probabilistically applied to the receiver's state, do not commute with every unitary operator. The class of simulable channels was expanded using a generalisation of the standard teleportation protocol, however this protocol is still not capable of simulating all channels~\cite{cope_simulation_2017}.

In Ref.~\cite{ishizaka_asymptotic_2008,ishizaka_quantum_2009}, Ishizaka and Hiroshima introduced a new teleportation protocol, called port-based teleportation (PBT). We consider the qubit version of this protocol. In the protocol, the sender and receiver each hold part of a resource state. Each qubit held by the receiver corresponds to a qubit held by the sender, and this shared two-qubit state is referred to as a port. In the standard case introduced by Ishizaka and Hiroshima, each port is an identical Bell pair. Then, a joint measurement is carried out on the sender's states and the qubit to be teleported; the result of this measurement is transmitted to the receiver, and based on this result, the receiver selects one of the ports and traces out the others. This measurement is chosen to be the square-root measurement, which projects the qubit to be teleported and one of the sender's resource qubits onto a Bell pair. The square-root measurement is optimal, in terms of entanglement fidelity, for PBT with a maximally entangled resource state, in both the qubit~\cite{ishizaka_asymptotic_2008} and qudit cases~\cite{mozrzymas_optimal_2018,leditzky_optimality_2020}. For a finite number of ports, $N$, the input-output channel from the PBT protocol is a depolarising channel whose diamond norm from the identity channel is exactly known~\cite{pirandola_fundamental_2019} and decreases to zero in the limit of $N \rightarrow \infty$.

In a more general setting, one can replace the original Bell pairs of the PBT protocol with any two-qubit state, and we may even allow entanglement between the ports. Here we investigate this general case, deriving the Choi matrix of the resulting PBT channel in terms of a multi-qubit resource state chosen for its ports. More precisely, we derive an explicit expression for the Choi matrix characterising the qubit channel given by enacting PBT using a given resource state and the square-root measurement (parametrised by the resource state). We make an assumption about the symmetry of the resource state under exchange of the labels of the ports, and show that resource states that fulfil this assumption can simulate any channel. We also show how this Choi matrix can be converted into the alternative channel representation of Kraus operators. We then find explicit expressions for the Kraus operators characterising the channel mapping from the resource state to the Choi matrix of the simulated qubit channel. These expressions characterise the PBT protocol itself, and can be used for optimising the finite-port simulation of a target channel over the convex set of resource states.

As an example of how the formulae can be applied, for two ports we give simple expressions for the Choi matrix of the simulated qubit channel. We also study families of resource states and, in particular, we define what we call ``Choi resources'', namely states made by $N$ copies of a generic state with a maximally mixed marginal. Via the Choi-Jamio{\l}kowski isomorphism \cite{choi_completely_1975,jamiolkowski_linear_1972}, each of these states is in one-to-one correspondence with a quantum channel, but not necessarily the channel we are interested in simulating. As an application, we simulate the amplitude damping (AD) channel using various Choi resources and show that a better simulation of the channel can be achieved using a Choi resource that corresponds to a possibly different AD channel. We find that the diamond norm (quantifying the quality of the simulation) can be found analytically at two different damping probabilities. Finally, we also investigate the simulation performance that is achievable with another family of resource states, with tensor-product structure, and such that they cannot be expressed as $N$ copies of the Choi matrix of some channel. At low $N$, this type of resource state is better at simulating an AD channel with low damping than any resource expressable as $N$ copies of the Choi matrix of an AD channel.

Our manuscript is structured as follows. In Sec.~\ref{secII}, we compute the expression of Choi matrix of the qubit channel simulated by PBT with an arbitrary multi-qubit resource state. In Sec.~\ref{secIII} we briefly discuss how to derive the Kraus operators of the simulated channel. Then, in Sec.~\ref{secIV}, we characterise the PBT map from the resource state to the Choi matrix of the simulated channel. In Sec.~\ref{secV}, we show the example of two-port PBT, in Sec.~\ref{secVI}, we use our expressions to calculate the depolarisation probability for PBT with a maximally entangled resource, and in Sec.~\ref{secVII}, we present the various improved simulations of the AD channel. Sec.~\ref{secVIII} is for conclusions.

\section{Calculating the Choi matrix for qubit PBT}\label{secII}
We consider an $N$-port qubit PBT protocol. We call the sender's part of the resource state the $A$ modes and the receiver's part of the resource state the $B$ modes. In order to characterise the channel simulated by PBT using a given resource state, we calculate the Choi matrix for that channel. To do so, we consider a maximally entangled 2-mode state, $|C_0C_1\rangle=\frac{1}{\sqrt{2}}\left(|00\rangle+|11\rangle\right)$. $C_0$ denotes the idler mode and $C_1$ denotes the teleported mode. The measurement consists of a POVM described by the operators $\hat{O}_i=\Pi_{i,AC_1}\otimes\mathbf{1}_{B C_0}$, where $i=1,\dots,N$. We consider the case in which the $\Pi_i$s describe a square-root measurement.  Given a certain measurement result $i$, Bob assumes that the state is teleported to the $i$-th mode $B_i$  and discards all the other ports via a partial trace applied to all $B_j$ with $j\neq i$, all the $A$ modes and $C_1$.

We assume that each port is symmetric under permutation of labels, i.e. that a swap operation that swaps both ports $A_i$ and $A_j$ and ports $B_i$ and $B_j$ does not change the density matrix of the resource state. This does not mean that the ports have to be independent of each other; it is still possible for the $A$ modes (or the $B$ modes, or both) to have some entanglement with each other. Consequently, all measurement outcomes are equally likely and all outcomes result in the same channel for the teleported state. We can therefore assume that the state is teleported to the first $B$ port without loss of generality, and so only consider one operator. We can justify this assumption as it simple to show that, for any non-symmetric resource state $\phi$, there exists a symmetric resource state $\phi^{\mathrm{sym}}$ that gives precisely the same channel \cite{banchi_convex_2020}.

Defining $\mathcal{P}_{\pi}$ as the qubit channel resulting from PBT using the program state $\pi$, we write
\begin{align}
\mathcal{P}_{\pi_{A B}}(\rho_{C_1})=\sum_{i=1}^N \Tr_{A \bar{B_i} C_1} \left[ \left(\sqrt{\Pi_i}_{A C_1}\otimes\mathbf{1}_{B}\right) \left(\pi_{A B}\otimes \rho_{C_1} \right) \left(\sqrt{\Pi_i}_{A C_1}\otimes\mathbf{1}_{B}\right)^{\dag} \right],
\end{align}
where $B_i$ is the port to which the state is teleported, $\bar{B_i}$ denotes all ports except for $B_i$ and $\Pi_i$ is the measurement operator applied to teleport the state to port $i$. Applying the symmetry condition, each value of $i$ gives the same output state, so we can carry out the sum and write
\begin{align}
\mathcal{P}_{\pi_{A B}}(\rho_{C_1})= N \Tr_{A \bar{B_1} C_1} \left[ \left(\sqrt{\Pi_1}_{A C_1}\otimes\mathbf{1}_{B}\right) \left(\pi_{A B}\otimes \rho_{C_1} \right) \left(\sqrt{\Pi_1}_{A C_1}\otimes\mathbf{1}_{B}\right)^{\dag} \right].
\end{align}
The Choi matrix of this channel is then given by
\begin{align}
\mathbf{1}_{C_0}\otimes\mathcal{P}_{\pi_{A B}}(\left|\Phi\right\rangle\left\langle\Phi\right|_{C_0 C_1}),
\end{align}
where $\left|\Phi\right\rangle=\frac{1}{\sqrt{2}}\left(|00\rangle+|11\rangle\right)$ is a Bell state.

For simplicity, let us initially consider what happens to a teleported arbitrary state $\rho_{C_1}$ (i.e. temporarily ignore the idler mode). Using the fact that the operator enacts the identity on the $B$ modes, we can take the trace on the $\bar{B}$ modes prior to the action of the operator. This allows us the simplification
\begin{align}
\mathcal{P}_{\pi_{A B}}(\rho_{C_1}) = N\Tr_{A C_1} \left[ \left(\sqrt{\Pi_1}_{A C_1}\otimes\mathbf{1}_{B_1}\right) \Tr_{\bar{B_1}}\left[\pi_{A B}\otimes \rho_{C_1} \right] \left(\sqrt{\Pi_1}_{A C_1}\otimes\mathbf{1}_{B_1}\right)^{\dag} \right].
\end{align}

We denote the matrix representation of $\mathcal{P}_{\pi_{A B}}(\rho_{C_1})$ as $V_{\mathrm{out}}$. We can then write
\begin{align}
V_{\mathrm{out}}&=\begin{pmatrix}
V_{\mathrm{out}}^{00} &V_{\mathrm{out}}^{01}\\
V_{\mathrm{out}}^{10} &V_{\mathrm{out}}^{11}
\end{pmatrix},\\
\begin{split}
V_{\mathrm{out}}^{ij}&=\left\langle i\middle|\mathcal{P}_{\pi_{A B}}(\rho_{C_1})\middle|j \right\rangle\\
&=N\left\langle i\middle|\Tr_{A C_1} \left[ \left(\sqrt{\Pi_1}_{A C_1}\otimes\mathbf{1}_{B_1}\right) \Tr_{\bar{B_1}}\left[\pi_{A B}\otimes \rho_{C_1} \right] \left(\sqrt{\Pi_1}_{A C_1}\otimes\mathbf{1}_{B_1}\right)^{\dag} \right]\middle|j\right\rangle.
\end{split}
\end{align}
Again using the fact that we enact the identity on the $B$ modes, we can take the contraction over the mode $B_1$ within the operation, arriving at
\begin{align}
\begin{split}
V_{\mathrm{out}}^{ij}&=N\Tr \left[ \sqrt{\Pi_1}_{A C_1} \left\langle i\middle|\Tr_{\bar{B_1}}\left[\pi_{A B}\otimes \rho_{C_1} \right]\middle|j\right\rangle \sqrt{\Pi_1}_{A C_1}^{\dag} \right]\\
&=N\Tr \left[ \Pi_1 \left\langle i\middle|\Tr_{\bar{B_1}}\left[\pi_{A B}\otimes \rho_{C_1} \right]\middle|j\right\rangle \right],
\end{split}
\end{align}
where we have used the cyclic invariance of the trace and the fact that $\Pi_1$ is a hermitian operator. In the second line and henceforth, we neglect the subscripts on $\Pi_1$. We now define $R^{i+1,j+1}=\left\langle i|_{B_1} \Tr_{\bar{B_1}}[\pi_{AB}] |j\right\rangle_{B_1}$ (the $+1$ is so that the labels run from 1 to 2 rather than from 0 to 1). Using this, we can simplify the expression for $V_{\mathrm{out}}$ to
\begin{align}
V_{\mathrm{out}}&=N\begin{pmatrix}
\Tr [\Pi_1(R^{11}\otimes\rho_{C_1})] &\Tr [\Pi_1(R^{12}\otimes\rho_{C_1})]\\
\Tr [\Pi_1(R^{21}\otimes\rho_{C_1})] &\Tr [\Pi_1(R^{22}\otimes\rho_{C_1})]
\end{pmatrix}.
\end{align}
Returning to considering the Choi matrix, $C$, we can use this simplification to write
\begin{align}
C=\frac{N}{2}\begin{pmatrix}
\chi_{00}^{11} &\chi_{00}^{12} &\chi_{01}^{11} &\chi_{01}^{12}\\
\chi_{00}^{21} &\chi_{00}^{22} &\chi_{01}^{21} &\chi_{01}^{22}\\
\chi_{10}^{11} &\chi_{10}^{12} &\chi_{11}^{11} &\chi_{11}^{12}\\
\chi_{10}^{21} &\chi_{10}^{22} &\chi_{11}^{21} &\chi_{11}^{22}
\end{pmatrix},\label{eq:Choi}\\
\chi_{mn}^{ij}=\Tr [\Pi_1(R^{ij}\otimes\left\lvert m\right\rangle\left\langle n\right\rvert_{C_1})].
\end{align}
It is worth noting that the Choi matrix is a valid density matrix, so we need only find expressions for the terms on or above the main diagonal. It is also worth noting that $R^{11}$ and $R^{22}$ are (unnormalised) density matrices, whilst $R^{12}$ and $R^{21}$ are not, in general.

Let us now consider the structure of the measurement $\Pi_1$, in a similar way to the analysis in \cite{ishizaka_quantum_2009}. $\Pi_1$ is a square-root measurement and can be linearly decomposed as $\Pi_1=\rho^{-\frac{1}{2}}\sigma_1\rho^{-\frac{1}{2}}+\frac{1}{N}(\mathbf{1}-\rho^{-\frac{1}{2}}\rho\rho^{-\frac{1}{2}})$, where $\sigma_i$ is the projector onto the Bell pair $\frac{1}{\sqrt{2}}(\left|01\right>-\left|10\right>)$ between qubit $C$ and the $i^{th}$ qubit in the sender's resource state (note that it is a different Bell pair from $\left|\Phi\right\rangle$, the Bell pair we used to define the Choi matrix) and $\rho=\sum_{i=1}^{N}\sigma_i$, as defined in \cite{ishizaka_quantum_2009}. Note that the powers of $\rho$ are taken over its support. Let us call the first term in this linear decomposition $M_1$ and call the second term $M_2$; we then have $\Pi_1=M_1+M_2$. Ishizaka and Hiroshima found that the eigenvalues of $\rho$ take one of two possible forms: $\lambda^-_j=\frac{1}{2}\left(\frac{N}{2}-j\right)$ or $\lambda^+_j=\frac{1}{2}\left(\frac{N}{2}+j+1\right)$ (these expressions differ slightly from those given in \cite{ishizaka_quantum_2009}, using a pre-factor of $\frac{1}{2}$ rather than $\frac{1}{2^N}$; this is purely due to defining $\sigma_i$ slightly differently). The two types of eigenvalue correspond to two types of eigenvector:
\begin{align}
&\left|\Psi(\lambda^{\mp}_j,m,\alpha)\right>=\Xi^{\pm -}(j,m+\frac{1}{2})\left|\Phi^{[N]}(j,m+\frac{1}{2},\alpha)\right>_A\left|0\right>_C+\Xi^{\pm +}(j,m-\frac{1}{2})\left|\Phi^{[N]}(j,m-\frac{1}{2},\alpha)\right>_A\left|1\right>_C,\\
\begin{split}
&\Xi^{+ +}(j,m)=\left<j,m,\frac{1}{2},\frac{1}{2}\middle|j+\frac{1}{2},m+\frac{1}{2}\right>,
\Xi^{+ -}(j,m)=\left<j,m,\frac{1}{2},-\frac{1}{2}\middle|j+\frac{1}{2},m-\frac{1}{2}\right>,\\
&\Xi^{- +}(j,m)=\left<j,m,\frac{1}{2},\frac{1}{2}\middle|j-\frac{1}{2},m+\frac{1}{2}\right>,
\Xi^{- -}(j,m)=\left<j,m,\frac{1}{2},-\frac{1}{2}\middle|j-\frac{1}{2},m-\frac{1}{2}\right>,
\end{split}
\end{align}
where $\Xi^{\pm \pm}(j,m)$ represents a Clebsch-Gordan coefficient, with the first superscripted sign determining whether $j$ increases or decreases by $\frac{1}{2}$ and the second superscripted sign determining whether $m$ increases or decreases by $\frac{1}{2}$. Note that $\langle{j,m,1/2,\pm1/2}\vert{J,M}\rangle=0$ if $|M|>J$ or $m\pm1/2\neq M$.

Ishizaka and Hiroshima treat the qubits as spins and hence treat the state $AC$ as a combination of an $N$-spin system and a spin singlet; $\left|\Phi^{[N]}(\lambda^{\mp}_j,m,\alpha)\right>$ then gives the orthogonal basis vectors of an $N$-spin system. $j$ corresponds to the magnitude of the spin of the resource state; this is a positive (half-)integer with minimum value 0 ($\frac{1}{2}$) when $N$ is even (odd). We call the magnitude of the total spin (of the $A$ and $C$ modes) $s$; $s$ has a maximum value of $\frac{N+1}{2}$, which occurs when every spin is aligned (all qubits in $AC$ are 0 or all are 1). $m$ corresponds to the spin of the total system in the z-direction. For fixed $s$, $m$ runs from $-s$ to $s$. The eigenvectors with eigenvalues $\lambda^-_j$ correspond to those states in which the total spin magnitude of the system $AC$ is the sum of the spin magnitudes of the systems $A$ and $C$ (i.e. the $A$ qubits have total spin $j$, the $C$ qubit has total spin $\frac{1}{2}$, so the system $AC$ has total spin $j+\frac{1}{2}$) and the eigenvectors with eigenvalues $\lambda^+_j$ correspond to states in which the spins subtract (i.e. the $A$ qubits have total spin $j$, the $C$ qubit has total spin $\frac{1}{2}$, so the system $AC$ has total spin $j-\frac{1}{2}$). Consequently, for fixed $s$, we have eigenvalues $\lambda^-_j$ with $j$ taking values up to $s-\frac{1}{2}$ and eigenvalues $\lambda^+_j$ with $j$ taking values up to $s+\frac{1}{2}$ (we also cannot have $\lambda^+_0$, since this would require the $A$ qubits to have negative total spin). For some values of $j$, multiple states $\left|\Phi^{[N]}(\lambda^{\mp}_j,m)\right>$ exist (i.e. $j$ and $m$ do not uniquely define a basis vector); in this case, we label the different states with $\alpha$, which runs from 1 to the degeneracy of the $j$-value, $g(N,j)$, (which depends only on $N$ and $j$, not on $m$).

Ishizaka and Hiroshima then divide the vectors in the $N$-spin basis into two types, based on how they are constructed from the $(N-1)$-spin basis; these are labelled $\left|\Phi^{[N]}_{I}(j,m,\alpha)\right>$ and $\left|\Phi^{[N]}_{II}(j,m,\alpha)\right>$. The eigenvectors of $\rho$ constructed using these basis vectors are then labelled $\left|\Psi_{I}(\lambda^{\mp}_j,m,\alpha)\right>$ and $\left|\Psi_{II}(\lambda^{\mp}_j,m,\alpha)\right>$. This categorisation is useful, because we can express $\rho$ and $M_1$ in terms of these vectors. The $N$-spin vectors are constructed as
\begin{align}
\begin{split}
\left|\Phi^{[N]}_{I}(j,m,\alpha)\right>&=\Xi^{- -}(j+\frac{1}{2},m+\frac{1}{2})\left|\Phi^{[N-1]}(j+\frac{1}{2},m+\frac{1}{2},\alpha)\right>_{\bar{A}}\left|0\right>_{A_1}\\
&+\Xi^{- +}(j+\frac{1}{2},m-\frac{1}{2})\left|\Phi^{[N-1]}(j+\frac{1}{2},m-\frac{1}{2},\alpha)\right>_{\bar{A}}\left|1\right>_{A_1},\label{eq:phi1}
\end{split}\\
\begin{split}
\left|\Phi^{[N]}_{II}(j,m,\alpha)\right>&=\Xi^{+ -}(j-\frac{1}{2},m+\frac{1}{2})\left|\Phi^{[N-1]}(j-\frac{1}{2},m+\frac{1}{2},\alpha)\right>_{\bar{A}}\left|0\right>_{A_1}\\
&+\Xi^{+ +}(j-\frac{1}{2},m-\frac{1}{2})\left|\Phi^{[N-1]}(j-\frac{1}{2},m-\frac{1}{2},\alpha)\right>_{\bar{A}}\left|1\right>_{A_1},\label{eq:phi2}
\end{split}
\end{align}
and the eigenvectors of $\rho$ are constructed as
\begin{align}
\begin{split}
\left|\Psi_{I}(\lambda^{\mp}_j,m,\alpha)\right>&=\Xi^{- -}(j+\frac{1}{2},m+1)\Xi^{\pm -}(j,m+\frac{1}{2})\left|\Phi^{[N-1]}(j+\frac{1}{2},m+1,\alpha)\right>_{\bar{A}}\left|00\right>_{A_{1}C}\\
&+\Xi^{- +}(j+\frac{1}{2},m)\Xi^{\pm -}(j,m+\frac{1}{2})\left|\Phi^{[N-1]}(j+\frac{1}{2},m,\alpha)\right>_{\bar{A}}\left|10\right>_{A_{1}C}\\
&+\Xi^{- -}(j+\frac{1}{2},m)\Xi^{\pm +}(j,m-\frac{1}{2})\left|\Phi^{[N-1]}(j+\frac{1}{2},m,\alpha)\right>_{\bar{A}}\left|01\right>_{A_{1}C}\\
&+\Xi^{- +}(j+\frac{1}{2},m-1)\Xi^{\pm +}(j,m-\frac{1}{2})\left|\Phi^{[N-1]}(j+\frac{1}{2},m-1,\alpha)\right>_{\bar{A}}\left|11\right>_{A_{1}C},\label{eq:psi1}
\end{split}
\end{align}
\begin{align}
\begin{split}
\left|\Psi_{II}(\lambda^{\mp}_j,m,\alpha)\right>&=\Xi^{+ -}(j-\frac{1}{2},m+1)\Xi^{\pm -}(j,m+\frac{1}{2})\left|\Phi^{[N-1]}(j-\frac{1}{2},m+1,\alpha)\right>_{\bar{A}}\left|00\right>_{A_{1}C}\\
&+\Xi^{+ +}(j-\frac{1}{2},m)\Xi^{\pm -}(j,m+\frac{1}{2})\left|\Phi^{[N-1]}(j-\frac{1}{2},m,\alpha)\right>_{\bar{A}}\left|10\right>_{A_{1}C}\\
&+\Xi^{+ -}(j-\frac{1}{2},m)\Xi^{\pm +}(j,m-\frac{1}{2})\left|\Phi^{[N-1]}(j-\frac{1}{2},m,\alpha)\right>_{\bar{A}}\left|01\right>_{A_{1}C}\\
&+\Xi^{+ +}(j-\frac{1}{2},m-1)\Xi^{\pm +}(j,m-\frac{1}{2})\left|\Phi^{[N-1]}(j-\frac{1}{2},m-1,\alpha)\right>_{\bar{A}}\left|11\right>_{A_{1}C}.\label{eq:psi2}
\end{split}
\end{align}
These explicit expressions will be useful later.

First, we write $\rho$ as a sum of projectors,
\begin{align}
\begin{split}
\rho=&\sum_{s=s_{\mathrm{min}}}^{\frac{N+1}{2}}\left[ \lambda^-_{s-\frac{1}{2}}\sum_{m=-s}^s\sum_\alpha \left(\left|\Psi_{I}(\lambda^-_{s-\frac{1}{2}},m,\alpha)\right>\left<\Psi_{I}(\lambda^-_{s-\frac{1}{2}},m,\alpha)\right| + \left|\Psi_{II}(\lambda^-_{s-\frac{1}{2}},m,\alpha)\right>\left<\Psi_{II}(\lambda^-_{s-\frac{1}{2}},m,\alpha)\right| \right)\right.\\
&\left.+\lambda^+_{s+\frac{1}{2}}\sum_{m=-s}^s\sum_\alpha \left(\left|\Psi_{I}(\lambda^+_{s+\frac{1}{2}},m,\alpha)\right>\left<\Psi_{I}(\lambda^+_{s+\frac{1}{2}},m,\alpha)\right| + \left|\Psi_{II}(\lambda^+_{s+\frac{1}{2}},m,\alpha)\right>\left<\Psi_{II}(\lambda^+_{s+\frac{1}{2}},m,\alpha)\right| \right) \right].
\end{split}
\end{align}
We then write $\rho^{-\frac{1}{2}}$ in the same way, getting
\begin{align}
\begin{split}
\rho^{-\frac{1}{2}}=&\sum_{s=s_{\mathrm{min}}}^{\frac{N+1}{2}}\left[ (\lambda^-_{s-\frac{1}{2}})^{-\frac{1}{2}}\sum_{m=-s}^s\sum_\alpha \left(\left|\Psi_{I}(\lambda^-_{s-\frac{1}{2}},m,\alpha)\right>\left<\Psi_{I}(\lambda^-_{s-\frac{1}{2}},m,\alpha)\right| + \left|\Psi_{II}(\lambda^-_{s-\frac{1}{2}},m,\alpha)\right>\left<\Psi_{II}(\lambda^-_{s-\frac{1}{2}},m,\alpha)\right| \right)\right.\\
&\left.+(\lambda^+_{s+\frac{1}{2}})^{-\frac{1}{2}}\sum_{m=-s}^s\sum_\alpha \left(\left|\Psi_{I}(\lambda^+_{s+\frac{1}{2}},m,\alpha)\right>\left<\Psi_{I}(\lambda^+_{s+\frac{1}{2}},m,\alpha)\right| + \left|\Psi_{II}(\lambda^+_{s+\frac{1}{2}},m,\alpha)\right>\left<\Psi_{II}(\lambda^+_{s+\frac{1}{2}},m,\alpha)\right| \right) \right].
\end{split}
\end{align}
The above expression is taken only over the support of $\rho$; some of the eigenvectors have an eigenvalue of 0, and we leave these out of the sum. From the form of the eigenvalues, we can see that they are all positive definite, except for in the case where $j=\frac{N}{2}$, and that $\lambda^-_{\frac{N}{2}}=0$. The corresponding eigenvectors, $\left|\Psi_{II}(\lambda^-_{\frac{N}{2}},m,\alpha)\right>$, define the vector space that is not part of the support of $\rho$ and hence the sum of the corresponding projectors gives us $M_2$ (since $\rho^{-\frac{1}{2}}\rho\rho^{-\frac{1}{2}}$ is the identity over the support of $\rho$). Note that there is no $\left|\Psi_{I}(\lambda^-_{\frac{N}{2}},m,\alpha)\right>$ vector, since this would require basis vectors of the $(N-1)$-spin subsystem with $j=\frac{N+1}{2}$ to exist. We can write the expression for $M_2$,
\begin{align}
M_2=\frac{1}{N}\sum_{m=-\frac{N+1}{2}}^{\frac{N+1}{2}}\sum_\alpha \left|\Psi_{II}(\lambda^-_{\frac{N}{2}},m,\alpha)\right>\left<\Psi_{II}(\lambda^-_{\frac{N}{2}},m,\alpha)\right|.\label{eq:M2}
\end{align}

We now want to find the form of $M_1=\rho^{-\frac{1}{2}}\sigma_1\rho^{-\frac{1}{2}}$. We express $\sigma_1$ as
\begin{align}
\sigma_1=\frac{1}{2}(\left|01\right>-\left|10\right>)(\left<01\right|-\left<10\right|)_{A_1C}\otimes\sum_{j=j_{\mathrm{min}}}^{\frac{N-1}{2}}\sum_{m=-j}^{j}\sum_{\alpha}\left| \Phi^{[N-1]}(j,m,\alpha)\right>\left<\Phi^{[N-1]}(j,m,\alpha)\right|_{\bar{A}}.
\end{align}
We then want to find $\frac{1}{\sqrt{2}}(\left<01\right|-\left<10\right|)_{A_{1}C}\left|\Psi_{I}(\lambda^{\mp}_{s\mp\frac{1}{2}},m,\alpha)\right>_{AC}$ and $\frac{1}{\sqrt{2}}(\left<01\right|-\left<10\right|)_{A_{1}C}\left|\Psi_{II}(\lambda^{\mp}_{s\mp\frac{1}{2}},m,\alpha)\right>_{AC}$; these will allow us to calculate $\rho^{-\frac{1}{2}}\sigma_1\rho^{-\frac{1}{2}}$. Ishizaka and Hiroshima calculated these using the expressions in Eqs.~(\ref{eq:psi1}) and (\ref{eq:psi2}) (and the explicit form of the Clebsch-Gordan coefficients), finding
\begin{align}
&\frac{1}{\sqrt{2}}(\left<01\right|-\left<10\right|)_{A_{1}C}\left|\Psi_{I}(\lambda^{-}_{s-\frac{1}{2}},m,\alpha)\right>_{AC}=\sqrt{\frac{s}{2s+1}}\left|\Phi^{[N-1]}(s,m,\alpha)\right>_{\bar{A}},\label{eq:comb1}\\
&\frac{1}{\sqrt{2}}(\left<01\right|-\left<10\right|)_{A_{1}C}\left|\Psi_{I}(\lambda^{+}_{s+\frac{1}{2}},m,\alpha)\right>_{AC}=0,\\
&\frac{1}{\sqrt{2}}(\left<01\right|-\left<10\right|)_{A_{1}C}\left|\Psi_{II}(\lambda^{-}_{s-\frac{1}{2}},m,\alpha)\right>_{AC}=0,\\
&\frac{1}{\sqrt{2}}(\left<01\right|-\left<10\right|)_{A_{1}C}\left|\Psi_{II}(\lambda^{+}_{s+\frac{1}{2}},m,\alpha)\right>_{AC}=-\sqrt{\frac{s+1}{2s+1}}\left|\Phi^{[N-1]}(s,m,\alpha)\right>_{\bar{A}}.\label{eq:comb2}
\end{align}
Combining our expressions for $\rho^{-\frac{1}{2}}$ and $\sigma_1$, and Eqs.~(\ref{eq:comb1}) to (\ref{eq:comb2}), we find that $M_1$ takes the form
\begin{align}
\begin{split}
M_1=&\sum_{s=s_{\mathrm{min}}}^{\frac{N-1}{2}}\sum_{m=-s}^s\sum_\alpha\left[ (\lambda^-_{s-\frac{1}{2}})^{-1}\frac{s}{2s+1} \left|\Psi_{I}(\lambda^-_{s-\frac{1}{2}},m,\alpha)\right>\left<\Psi_{I}(\lambda^-_{s-\frac{1}{2}},m,\alpha)\right| \right.\\
&-(\lambda^-_{s-\frac{1}{2}}\lambda^+_{s+\frac{1}{2}})^{-\frac{1}{2}}\frac{\sqrt{s(s+1)}}{2s+1} \left( \left|\Psi_{I}(\lambda^-_{s-\frac{1}{2}},m,\alpha)\right>\left<\Psi_{II}(\lambda^+_{s+\frac{1}{2}},m,\alpha)\right|+\left|\Psi_{II}(\lambda^+_{s+\frac{1}{2}},m,\alpha)\right>\left<\Psi_{I}(\lambda^-_{s-\frac{1}{2}},m,\alpha)\right| \right)\\
&\left.+(\lambda^+_{s+\frac{1}{2}})^{-1}\frac{s+1}{2s+1} \left|\Psi_{II}(\lambda^+_{s+\frac{1}{2}},m,\alpha)\right>\left<\Psi_{II}(\lambda^+_{s+\frac{1}{2}},m,\alpha)\right| \right].\label{eq:M1}
\end{split}
\end{align}
We have summed $s$ from $s_{\mathrm{min}}$ to $\frac{N-1}{2}$, rather than to $\frac{N+1}{2}$, since $\lambda^-_{\frac{N}{2}}=0$ and the vector $\left|\Psi(\lambda^{+}_{\frac{N}{2}+1},m,\alpha)\right>$ does not exist.

We now calculate $\left<0\middle|\Psi_{I}(\lambda^{\mp}_{s\mp\frac{1}{2}},m,\alpha)\right>$, $\left<0\middle|\Psi_{II}(\lambda^{\mp}_{s\mp\frac{1}{2}},m,\alpha)\right>$, $\left<1\middle|\Psi_{I}(\lambda^{\mp}_{s\mp\frac{1}{2}},m,\alpha)\right>$ and $\left<1\middle|\Psi_{II}(\lambda^{\mp}_{s\mp\frac{1}{2}},m,\alpha)\right>$ (where the contraction is over the $C$ qubit). Using the expressions in Eqs.~(\ref{eq:psi1}) and (\ref{eq:psi2}), and finding the explicit form of the Clebsch-Gordan coefficients~\cite{ishizaka_quantum_2009}, we calculate
\begin{align}
&\left<0\middle|\Psi_{I(II)}(\lambda^{-}_{s-\frac{1}{2}},m,\alpha)\right>=\sqrt{\frac{1}{2}-\frac{m}{2s}}\left|\Phi^{[N]}_{I(II)}(s-\frac{1}{2},m+\frac{1}{2},\alpha)\right>_A,\label{eq:comb3}\\
&\left<1\middle|\Psi_{I(II)}(\lambda^{-}_{s-\frac{1}{2}},m,\alpha)\right>=\sqrt{\frac{1}{2}+\frac{m}{2s}}\left|\Phi^{[N]}_{I(II)}(s-\frac{1}{2},m-\frac{1}{2},\alpha)\right>_A,\\
&\left<0\middle|\Psi_{I(II)}(\lambda^{+}_{s+\frac{1}{2}},m,\alpha)\right>=\sqrt{\frac{1}{2}+\frac{m}{2(s+1)}}\left|\Phi^{[N]}_{I(II)}(s+\frac{1}{2},m+\frac{1}{2},\alpha)\right>_A,\\
&\left<1\middle|\Psi_{I(II)}(\lambda^{+}_{s+\frac{1}{2}},m,\alpha)\right>=-\sqrt{\frac{1}{2}-\frac{m}{2(s+1)}}\left|\Phi^{[N]}_{I(II)}(s+\frac{1}{2},m-\frac{1}{2},\alpha)\right>_A.\label{eq:comb4}
\end{align}

We now have enough to start calculating the components of the Choi matrix. As an example, let us consider the top-left component, $\chi_{00}^{11}$. We are given $R^{11}$, $R^{12}$ and $R^{22}$ as the specification of the resource state. Let us demand that these are given in the $N$-spin basis (the $\left|\Phi^{[N]}_{I(II)}(j,m,\alpha)\right>$ basis). In order to make it clear which components of the resource state we are referring to without choosing some specific matrix representation, we define the function $f^{11}_{I,I}$ such that $f^{11}_{I,I}(j_1,m_1,\alpha_1,j_2,m_2,\alpha_2)$ is the coefficient of $\left|\Phi^{[N]}_{I}(j_1,m_1,\alpha_1)\right>\left<\Phi^{[N]}_{I}(j_2,m_2,\alpha_2)\right|$ in $R^{11}$. We similarly define $f^{11}_{I,II}$, $f^{11}_{II,I}$ and $f^{11}_{II,II}$, and similar functions for $R^{12}$, $R^{21}$ and $R^{22}$. These functions are simply a way of specifying the resource state. Together, $R^{11}$, $R^{12}$ and $R^{22}$ give the resource state after tracing over all but one $B$ mode. With our assumption that the resource state is unchanged by a swap operation between two ports, this is sufficient to specify the resource state.

We then calculate contributions to the Choi matrix from $M_1$ and $M_2$, using the expressions in Eqs.~(\ref{eq:M1}) and (\ref{eq:M2}), and Eqs.~(\ref{eq:comb3}) to (\ref{eq:comb4}). Recall that $M_1$ acts on the support of $\rho$ and $M_2$ acts on the part of the resource state that is not on the support of $\rho$. The contribution to $\chi_{00}^{11}$ from $M_1$ is
\begin{align}
&\begin{aligned}
\Tr [M_1(R^{11}\otimes&\left|0\right>\left<0\right|_{C_1})]=\sum_{s=s_{\mathrm{min}}}^{\frac{N-1}{2}}\sum_{m=-s}^s\sum_\alpha\left[ q_{-}^2 f^{11}_{I,I}(s-\frac{1}{2},m+\frac{1}{2},\alpha,s-\frac{1}{2},m+\frac{1}{2},\alpha) \right.\\
&-q_{-}r_{+} \left(f^{11}_{I,II}(s-\frac{1}{2},m+\frac{1}{2},\alpha,s+\frac{1}{2},m+\frac{1}{2},\alpha)+f^{11}_{II,I}(s+\frac{1}{2},m+\frac{1}{2},\alpha,s-\frac{1}{2},m+\frac{1}{2},\alpha)\right)\\
&\left.+r_{+}^2 f^{11}_{II,II}(s+\frac{1}{2},m+\frac{1}{2},\alpha,s+\frac{1}{2},m+\frac{1}{2},\alpha) \right],
\end{aligned}\\
&q_{\pm}=\sqrt{\frac{2(s \pm m)}{(N+1-2s)(2s+1)}},\label{eq:q_def}\\
&r_{\pm}=\sqrt{\frac{2(s \pm m+1)}{(N+3+2s)(2s+1)}},\label{eq:r_def}
\end{align}
where we have used the explicit form of the eigenvalues. The contribution to $\chi_{00}^{11}$ from $M_2$ is
\begin{align}
\Tr [M_2&(R^{11}\otimes\left|0\right>\left<0\right|_{C_1})]=\frac{1}{N}\sum_{m=-\frac{N+1}{2}}^{\frac{N+1}{2}} (\frac{1}{2}-\frac{m}{N+1})f^{11}_{II,II}(\frac{N}{2},m+\frac{1}{2},1,\frac{N}{2},m+\frac{1}{2},1).
\end{align}
We do not need to sum over $\alpha$, since there is no degeneracy in the states we sum over. By adding these two contributions and multiplying by $\frac{N}{2}$ (as per Eq.~(\ref{eq:Choi})), we get the top-left component of the Choi matrix. We call this component $C^{11}$. Then,
\begin{align}
\begin{split}
C^{11}=&\frac{N}{2}\sum_{s=s_{\mathrm{min}}}^{\frac{N-1}{2}}\sum_{m=-s}^s\sum_\alpha\left[ q_{-}^2 f^{11}_{I,I}(s-\frac{1}{2},m+\frac{1}{2},\alpha,s-\frac{1}{2},m+\frac{1}{2},\alpha) \right.\\
&-q_{-}r_{+} \left(f^{11}_{I,II}(s-\frac{1}{2},m+\frac{1}{2},\alpha,s+\frac{1}{2},m+\frac{1}{2},\alpha)+f^{11}_{II,I}(s+\frac{1}{2},m+\frac{1}{2},\alpha,s-\frac{1}{2},m+\frac{1}{2},\alpha)\right)\\
&\left.+r_{+}^2 f^{11}_{II,II}(s+\frac{1}{2},m+\frac{1}{2},\alpha,s+\frac{1}{2},m+\frac{1}{2},\alpha) \right]+\frac{1}{2}\sum_{m=-\frac{N+1}{2}}^{\frac{N+1}{2}} (\frac{1}{2}-\frac{m}{N+1})f^{11}_{II,II}(\frac{N}{2},m+\frac{1}{2},1,\frac{N}{2},m+\frac{1}{2},1).
\end{split}
\end{align}
We can express this more succinctly by defining the functions
\begin{align}
g^{a}_{b}[-+-+](s,m)= \sum_\alpha f^{a}_{b}(s-\frac{1}{2},m+\frac{1}{2},\alpha,s-\frac{1}{2},m+\frac{1}{2},\alpha),
\end{align}
where the index $a$ could be ``11", ``12", ``21" or ``22" and the index $b$ could be ``$I,I$", ``$I,II$", ``$II,I$" or ``$II,II$". Equally, the signs given as arguments to the $g$ function can be changed (e.g. we could have ``++++" instead of ``-+-+"), and in this case the signs in the $f$ function change accordingly. We can then express $C^{11}$ as
\begin{align}
\begin{split}
C^{11}=&\frac{N}{2}\sum_{s=s_{\mathrm{min}}}^{\frac{N-1}{2}}\sum_{m=-s}^s\left[ q_{-}^2 g^{11}_{I,I}[-+-+](s,m) -q_{-}r_{+} \left(g^{11}_{I,II}[-+++](s,m) + g^{11}_{II,I}[++-+](s,m)\right)\right.\\
&\left.+r_{+}^2 g^{11}_{II,II}[++++](s,m) \right] + \frac{1}{2}\sum_{m=-\frac{N+1}{2}}^{\frac{N+1}{2}} (\frac{1}{2}-\frac{m}{N+1})g^{11}_{II,II}[-+-+](\frac{N+1}{2},m).
\end{split}\label{eq:c11}
\end{align}

To get the expressions for $C^{12}$ and $C^{22}$, we simply replace $g^{11}$ with $g^{12}$ and $g^{22}$ respectively in the expression for $C^{11}$. Equally, once we have the expression for $C^{13}$, we can get the expressions for $C^{14}$, $C^{23}$ and $C^{24}$ by replacing $g^{11}$ with $g^{12}$, $g^{21}$ and $g^{22}$ respectively in the expression for $C^{13}$. Similarly, starting from the expressions for $C^{33}$, we get the expressions for $C^{34}$ and $C^{44}$ by replacing $g^{11}$ with $g^{12}$ and $g^{22}$ respectively in the expression for $C^{33}$. Essentially, if we divide the Choi matrix into quarters, we only need one expression per block of four elements, and the other expressions only require trivial modifications. We also only need the expressions for the upper triangle of the Choi matrix, since the Choi matrix is a valid density matrix and so is hermitian. We give the expressions for $C^{13}$ and $C^{33}$ below:
\begin{align}
\begin{split}
C^{13}=&\frac{N}{2}\sum_{s=s_{\mathrm{min}}}^{\frac{N-1}{2}}\sum_{m=-s}^s\left[ q_{-}q_{+} g^{11}_{I,I}[-+--](s,m) +q_{-}r_{-} g^{11}_{I,II}[-++-](s,m) -q_{+}r_{+} g^{11}_{II,I}[++--](s,m)\right.\\
&\left. -r_{-}r_{+} g^{11}_{II,II}[+++-](s,m) \right]+\frac{1}{2}\sum_{m=-\frac{N+1}{2}}^{\frac{N+1}{2}} \sqrt{\frac{1}{4}-\left(\frac{m}{N+1}\right)^2}g^{11}_{II,II}[-+--](\frac{N+1}{2},m),
\end{split}\label{eq:c13}\\
\begin{split}
C^{33}=&\frac{N}{2}\sum_{s=s_{\mathrm{min}}}^{\frac{N-1}{2}}\sum_{m=-s}^s\left[ q_{+}^2 g^{11}_{I,I}[----](s,m) +q_{+}r_{-} \left(g^{11}_{I,II}[--+-](s,m)+g^{11}_{II,I}[+---](s,m)\right)\right.\\
&\left.+r_{-}^2 g^{11}_{II,II}[+-+-](s,m) \right]+\frac{1}{2}\sum_{m=-\frac{N+1}{2}}^{\frac{N+1}{2}} (\frac{1}{2}+\frac{m}{N+1})g^{11}_{II,II}[----](\frac{N+1}{2},m).
\end{split}\label{eq:c33}
\end{align}
These are, in fact, fairly simple expressions, although quite long when written in this form. If we impose constraints on the resource state, we can simplify the expressions.

We now have an analytical expression for the Choi matrix for any PBT qubit operation. The only assumption made is that all ports are identical. Any channel simulable via PBT can be simulated using a resource state of this type~\cite{christandl_asymptotic_2019}. See the supplementary material for a Mathematica implementation of these expressions~\cite{zenodo}.

To show how the Choi matrix, $C$, is constructed from the components given, we write the following, where * denotes the complex conjugate and where $C^{ij}(g^{11}\to g^{kl})$ means the expression for $C^{ij}$ with all instances of $g^{11}$ replaced with $g^{kl}$:
\begin{align}
C=\begin{pmatrix}
C^{11}(g^{11}) &C^{11}(g^{11} \to g^{12}) &C^{13}(g^{11}) &C^{13}(g^{11} \to g^{12})\\
C^{11}(g^{11} \to g^{12})^* &C^{11}(g^{11} \to g^{22}) &C^{13}(g^{11} \to g^{21}) &C^{13}(g^{11} \to g^{22})\\
C^{13}(g^{11})^* &C^{13}(g^{11} \to g^{21})^* &C^{33}(g^{11}) &C^{33}(g^{11} \to g^{12})\\
C^{13}(g^{11} \to g^{12})^* &C^{13}(g^{11} \to g^{22})^* &C^{33}(g^{11} \to g^{12})^* &C^{33}(g^{11} \to g^{22})
\end{pmatrix}.\label{eq:total_choi}
\end{align}

We may also wish to find the Kraus operators \cite{choi_completely_1975} of the qubit channel resulting from PBT using a given resource state. This is an alternative but equivalent channel representation to the Choi matrix. We may also wish to characterise the channel mapping from a given resource state to the output Choi matrix of the qubit channel. This channel takes a resource state as input and outputs the Choi matrix of the qubit channel resulting from PBT using that resource state. These Kraus operators are rectangular (the number of qubits in the output is less than the number in the input). They characterise the processor (i.e. the operation of carrying out a square-root measurement on the modes $AC_1$, followed by the selection of a $B$ port based on the measurement outcome).

\section{Converting from the Choi matrix to the Kraus operators of the qubit channel}\label{secIII}
The Choi matrix holds all information about the state, but we would like to also be able to express the channel as a set of Kraus operators \cite{choi_completely_1975}. We can do this using the following algorithm, starting from the Choi matrix $V$.

\begin{enumerate}
\item
Find the eigendecomposition of $V$, and write:
\begin{align}
V=\sum_{i=1}^4 \lambda_i \left|v_i'\right>\left<v_i'\right|.
\end{align}
\item
We then define $\left|v_i\right>=\sqrt{\lambda_i}\left|v_i'\right>$, so that we can write:
\begin{align}
V=\sum_{i=1}^4 \left|v_i\right>\left<v_i\right|.
\end{align}
\item
The (up to) four Kraus operators, labelled as $K_i$, are then written (in the canonical basis) as
\begin{align}
K_i=\begin{pmatrix}
\left< 00 | v_i \right> &\left< 10 | v_i \right>\\
\left< 01 | v_i \right> &\left< 11 | v_i \right>
\end{pmatrix}
\end{align}
\end{enumerate}

We can verify that if the Kraus operators constructed in this way are applied to a Bell state, we recover the initial Choi matrix. Numerically, this algorithm is simple to implement, since we are only finding the eigendecomposition of a 4 by 4 matrix.

\section{Finding the Kraus operators of the channel from the program state to the Choi matrix of the simulated qubit channel}\label{secIV}
We want to characterise the channel mapping from the (input) program state (with $2N$ qubits) to the (output) Choi matrix of the PBT channel (with 2 qubits). This is a characterisation of the PBT protocol itself (with the square-root measurement and a permutation-symmetric resource state). An implicit expression for this map is derived in \cite{banchi_convex_2020}, however here we derive explicit expressions.

Defining $\Lambda$ as the channel from the program state to the Choi matrix of the qubit channel, we can write
\begin{align}
\begin{split}
\Lambda(\pi)&=\sum_{i=1}^N \Tr_{A \bar{B_i} C_1} \left[ \left(\sqrt{\Pi_i}_{A C_1}\otimes\mathbf{1}_{B C_0}\right) \left(\pi_{A B}\otimes\left|\Phi_{C_0 C_1}\right>\left<\Phi_{C_0 C_1}\right| \right) \left(\sqrt{\Pi_i}_{A C_1}\otimes\mathbf{1}_{B C_0}\right)^{\dag} \right]\\
&=\sum_{ik} K_{ik} \pi K_{ik}^{\dag},
\end{split}
\end{align}
where $B_i$ is the port to which the state is teleported, $\Pi_i$ is the measurement operator applied to teleport the state to port i and
\begin{align}
K_{ik}=\left<e_k^{(i)}\middle|\sqrt{\Pi_i}_{A C_1}\otimes\mathbf{1}_{B C_0}\middle|\Phi_{C_0 C_1}\right>.
\end{align}
The $\left|e_k^{(i)}\right>$ are basis vectors on the systems $A\bar{B_i}C_1$ (the traced over systems).

First, let us apply the assumption of symmetry under exchange of labels. We can therefore replace $K_{ik}$ with $K_{k}=\sqrt{N}K_{1k}$. We can now calculate $\sqrt{\Pi_1}$, using the expressions in Eqs.~(\ref{eq:M1}) and (\ref{eq:M2}). From the fact that $M_1$ and $M_2$ have orthogonal supports, we can take the square roots of each separately. In fact, due to $M_1$ having no mixing between basis vectors with different $s$, $m$ or $\alpha$ values, we can treat each set of values $\{s,m,\alpha\}$ separately, and hence can write
\begin{align}
\sqrt{\Pi_1}=\sum_{sm\alpha} \sqrt{M_1^{sm\alpha}} + \sqrt{M_2},
\end{align}
where $M_1^{sm\alpha}$ is the contribution to $M_1$ from the two eigenvectors $\left|\Psi_{I}(\lambda^-_{s-\frac{1}{2}},m,\alpha)\right>$ and $\left|\Psi_{II}(\lambda^+_{s+\frac{1}{2}},m,\alpha)\right>$. Since $M_2$, as expressed in Eq.~(\ref{eq:M2}), is already diagonal, it is trivial to write
\begin{align}
\sqrt{M_2}=\frac{1}{\sqrt{N}}\sum_{m=-\frac{N+1}{2}}^{\frac{N+1}{2}} \left|\Psi_{II}(\lambda^-_{\frac{N}{2}},m)\right>\left<\Psi_{II}(\lambda^-_{\frac{N}{2}},m)\right|,
\end{align}
where we have removed the sum over $\alpha$, due to there being no degeneracy in the component eigenvectors.

We now want to find $\sqrt{M_1^{sm\alpha}}$, starting from
\begin{align}
\begin{split}
M_1^{sm\alpha}&=(\lambda^-_{s-\frac{1}{2}})^{-1}\frac{s}{2s+1} \left|\Psi_{I}(\lambda^-_{s-\frac{1}{2}},m,\alpha)\right>\left<\Psi_{I}(\lambda^-_{s-\frac{1}{2}},m,\alpha)\right|\\
&-(\lambda^-_{s-\frac{1}{2}}\lambda^+_{s+\frac{1}{2}})^{-\frac{1}{2}}\frac{\sqrt{s(s+1)}}{2s+1} \left( \left|\Psi_{I}(\lambda^-_{s-\frac{1}{2}},m,\alpha)\right>\left<\Psi_{II}(\lambda^+_{s+\frac{1}{2}},m,\alpha)\right|+\left|\Psi_{II}(\lambda^+_{s+\frac{1}{2}},m,\alpha)\right>\left<\Psi_{I}(\lambda^-_{s-\frac{1}{2}},m,\alpha)\right| \right)\\
&+(\lambda^+_{s+\frac{1}{2}})^{-1}\frac{s+1}{2s+1} \left|\Psi_{II}(\lambda^+_{s+\frac{1}{2}},m,\alpha)\right>\left<\Psi_{II}(\lambda^+_{s+\frac{1}{2}},m,\alpha)\right|.
\end{split}\label{eq:M1sma}
\end{align}
From the form of Eq.~(\ref{eq:M1sma}), we can see that $M_1^{sm\alpha}$ can be written as
\begin{align}
&M_1^{sm\alpha}=\left|\textrm{vec}^{sm\alpha}\right>\left<\textrm{vec}^{sm\alpha}\right|,\\
&\left|\textrm{vec}^{sm\alpha}\right>=\sqrt{(\lambda^-_{s-\frac{1}{2}})^{-1}\frac{s}{2s+1}}\left|\Psi_{I}(\lambda^-_{s-\frac{1}{2}},m,\alpha)\right> - \sqrt{(\lambda^+_{s+\frac{1}{2}})^{-1}\frac{s+1}{2s+1}}\left|\Psi_{II}(\lambda^+_{s+\frac{1}{2}},m,\alpha)\right>,
\end{align}
where it must be noted that $\left|\textrm{vec}^{sm\alpha}\right>$ is unnormalised. This means that $M_1^{sm\alpha}$ has only one non-zero eigenvalue, given by
\begin{equation}
\begin{split}
\textrm{eig}^{sm\alpha}&=(\lambda^-_{s-\frac{1}{2}})^{-1}\frac{s}{2s+1}+(\lambda^+_{s+\frac{1}{2}})^{-1}\frac{s+1}{2s+1}\\
&=\frac{4(N+1)}{(N+1-2s)(N+3+2s)}
\end{split}.
\end{equation}
Consequently, we can write
\begin{equation}
\sqrt{M_1^{sm\alpha}}=(\textrm{eig}^{sm\alpha})^{-\frac{1}{2}}\left|\textrm{vec}^{sm\alpha}\right>\left<\textrm{vec}^{sm\alpha}\right|.
\end{equation}

Combining our expressions for $M_1$ and $M_2$, we have
\begin{align}
\sqrt{\Pi_1}=\frac{1}{\sqrt{N}}\sum_m \left|\Psi_{II}(\lambda^-_{\frac{N}{2}},m)\right>\left<\Psi_{II}(\lambda^-_{\frac{N}{2}},m)\right| + \sum_{sm\alpha} \sqrt{\frac{(N+1-2s)(N+3+2s)}{4(N+1)}}\left|\textrm{vec}^{sm\alpha}\right>\left<\textrm{vec}^{sm\alpha}\right|.
\end{align}
We now express the basis vectors $\left|e_k^{(1)}\right>$ as
\begin{align}
\left|e_k\right>=\left|e_{k_1}\right>_{AC_1}\left|e_{k_2}\right>_{\bar{B}},
\end{align}
where $\bar{B}$ refers to the $B$ modes except for $B_1$. $\left|e_{k_1}\right>_{AC_1}$ are the $\left|\textnormal{vec}_2^{sm\alpha}\right>$ basis vectors (on the system $AC_1$) and the $\left|e_{k_2}\right>_{\bar{B}}$ are any choice of orthonormal basis vectors on the system $\bar{B}$. There are two types of Kraus operator, depending on whether $\left|e_{k_1}\right>_{AC_1}$ lies in the support of $M_1$ or of $M_2$. We will label these Kraus operators $K_k^1$ and $K_k^2$ respectively. Using Eqs.~(\ref{eq:comb3}) to (\ref{eq:comb4}), we find that the Kraus operators $K_k^2$ take the form
\begin{align}
K_k^2=\frac{1}{\sqrt{2}}\left( \sqrt{\frac{1}{2}-\frac{m}{N+1}} \left|0\right>_{C_0}\left<\Phi_{II}(\frac{N}{2},m+\frac{1}{2})\right|_{AC_1} + \sqrt{\frac{1}{2}+\frac{m}{N+1}} \left|1\right>_{C_0}\left<\Phi_{II}(\frac{N}{2},m-\frac{1}{2})\right|_{AC_1} \right)\left<e_{k_2}\right|_{\bar{B}}\otimes\mathbf{1}_{B_1},\label{eq:kraus2}
\end{align}
where the label $k$ determines the $m$ value and the choice of basis vector $\left|e_{k_2}\right>_{\bar{B}}$. We find that the Kraus operators $K_k^1$ take the form
\begin{align}
\begin{split}
K_k^1&=\sqrt{\frac{N}{2}}\left[ \left|0\right>_{C_0}\left( \sqrt{(\lambda^-_{s-\frac{1}{2}})^{-1}\frac{s}{2s+1}\left(\frac{1}{2}-\frac{m}{2s}\right)} \left<\Phi_{I}(s-\frac{1}{2},m+\frac{1}{2},\alpha)\right|\right.\right.\\
&\left.- \sqrt{(\lambda^+_{s+\frac{1}{2}})^{-1}\frac{s+1}{2s+1}\left(\frac{1}{2}+\frac{m}{2(s+1)}\right)} \left<\Phi_{II}(s+\frac{1}{2},m+\frac{1}{2},\alpha)\right| \right)_{AC_1}\\
&+ \left|1\right>_{C_0}\left( \sqrt{(\lambda^-_{s-\frac{1}{2}})^{-1}\frac{s}{2s+1}\left(\frac{1}{2}+\frac{m}{2s}\right)} \left<\Phi_{I}(s-\frac{1}{2},m-\frac{1}{2},\alpha)\right|\right.\\
&\left.\left.+ \sqrt{(\lambda^+_{s+\frac{1}{2}})^{-1}\frac{s+1}{2s+1}\left(\frac{1}{2}-\frac{m}{2(s+1)}\right)} \left<\Phi_{II}(s+\frac{1}{2},m-\frac{1}{2},\alpha)\right| \right)_{AC_1} \right]\left<e_{k_2}\right|_{\bar{B}}\otimes\mathbf{1}_{B_1},
\end{split}
\end{align}
where the label $k$ determines the values of $s$, $m$ and $\alpha$, and the choice of basis vector $\left|e_{k_2}\right>_{\bar{B}}$. We can simplify this expression, and so can write
\begin{align}
\begin{split}
K_k^1=&\sqrt{\frac{N}{2}}\left[ \left|0\right>_{C_0}\left( q_{-} \left<\Phi_{I}(s-\frac{1}{2},m+\frac{1}{2},\alpha)\right|- r_{+} \left<\Phi_{II}(s+\frac{1}{2},m+\frac{1}{2},\alpha)\right| \right)_{AC_1}\right.\\
&\left. +\left|1\right>_{C_0}\left( q_{+} \left<\Phi_{I}(s-\frac{1}{2},m-\frac{1}{2},\alpha)\right|+ r_{-} \left<\Phi_{II}(s+\frac{1}{2},m-\frac{1}{2},\alpha)\right| \right)_{AC_1} \right]\left<e_{k_2}\right|_{\bar{B}}\otimes\mathbf{1}_{B_1},
\end{split}\label{eq:kraus1}
\end{align}
where $q_{\pm}$ and $r_{\pm}$ are defined as per Eqs.~(\ref{eq:q_def}) and (\ref{eq:r_def}).

Note that the basis vectors $\left|e_{k_2}\right>_{\bar{B}}$ simply trace over the $\bar{B}$ system, i.e. for each Kraus operator, there are $2^N-1$ other Kraus operators that are identical up to a change in $k_2$. Hence, we can trace over the $\bar{B}$ modes of the resource state; in this case the Kraus operators of the channel from $\Tr_{\bar{B}}\left[\pi_{AB}\right]$ to the output Choi matrix are $K_k^1$ and $K_k^2$ without the vectors $\left|e_{k_2}\right>_{\bar{B}}$ (i.e. the labels $k$ determine only the values of $s$, $m$ and $\alpha$).

\section{Two port PBT}\label{secV}
As an example, suppose we only have two ports. Let us calculate the Choi matrix for this case. We again assume that the two ports are identical under exchange of labels. The reduced resource states $R^{11}$, $R^{12}$ and $R^{22}$ are then 4 by 4 matrices. We will write them in the basis: $\{\frac{1}{\sqrt{2}}(\left|10\right>-\left|01\right>),\left|00\right>,\frac{1}{\sqrt{2}}(\left|10\right>+\left|01\right>),\left|11\right>\}$. These are the vectors $\{\left|\Phi^{[2]}_{I}(0,0)\right>,\left|\Phi^{[2]}_{II}(1,-1)\right>,\left|\Phi^{[2]}_{II}(1,0)\right>,\left|\Phi^{[2]}_{II}(1,1)\right>\}$. Note that there are no degenerate $(j,m)$ combinations for two ports, so we do not need to specify the degeneracy, $\alpha$. We can therefore immediately remove the sum over $\alpha$. Since $s=\frac{1}{2}$ is the only value of $s$ for which either $\left|\Phi^{[2]}_{I}(s-\frac{1}{2},m)\right>$ or $\left|\Phi^{[2]}_{II}(s+\frac{1}{2},m)\right>$ exist, we do not need to sum over $s$ either, and simply set $s=\frac{1}{2}$. $R^{ij}$ takes the form
\begin{align}
R^{ij}=\begin{pmatrix}
f_{I,I}^{ij}(0,0,0,0) &f_{I,II}^{ij}(0,0,1,-1) &f_{I,II}^{ij}(0,0,1,0) &f_{I,II}^{ij}(0,0,1,1)\\
f_{II,I}^{ij}(1,-1,0,0) &f_{II,II}^{ij}(1,-1,1,-1) &f_{II,II}^{ij}(1,-1,1,0) &f_{II,II}^{ij}(1,-1,1,1)\\
f_{II,I}^{ij}(1,0,0,0) &f_{II,II}^{ij}(1,0,1,-1) &f_{II,II}^{ij}(1,0,1,0) &f_{II,II}^{ij}(1,0,1,1)\\
f_{II,I}^{ij}(1,1,0,0) &f_{II,II}^{ij}(1,1,1,-1) &f_{II,II}^{ij}(1,1,1,0) &f_{II,II}^{ij}(1,1,1,1)
\end{pmatrix},
\end{align}
where we have excluded $\alpha$ from the arguments of $f$. We again note that $R^{11}$, $R^{12}$, $R^{21}$ and $R^{22}$ are derived from the density matrix of the full resource state by taking the trace over all $B$ modes except for the first $B$ mode. In the two mode case, they can be written as
\begin{align}
R^{11}=\left<0\right|_{B1}\Tr_{B2}\left(R\right)\left|0\right>_{B1},\\
R^{12}=\left<0\right|_{B1}\Tr_{B2}\left(R\right)\left|1\right>_{B1},\\
R^{21}=\left<1\right|_{B1}\Tr_{B2}\left(R\right)\left|0\right>_{B1},\\
R^{22}=\left<1\right|_{B1}\Tr_{B2}\left(R\right)\left|1\right>_{B1}.
\end{align}
The expression for $C^{11}$, in Eq.~(\ref{eq:c11}), now reduces to
\begin{align}
C^{11}=\frac{1}{2} \Tr \left[R^{11}\right] - \frac{1}{2\sqrt{3}}\left( f_{I,II}^{11}(0,0,1,0) + f_{II,I}^{11}(1,0,0,0) \right),\label{eq:c11_2qubit}
\end{align}
where we have used
\begin{align}
\begin{split}
\Tr \left[R^{11}\right]&=f_{I,I}^{11}(0,0,0,0) + f_{II,II}^{11}(1,-1,1,-1) + f_{II,II}^{11}(1,0,1,0) + f_{II,II}^{11}(1,1,1,1)\\
&=\Tr\left[\left<0\right|_{B1}R\left|0\right>_{B1}\right].
\end{split}
\end{align}
The expressions for $C^{13}$ and $C^{33}$, in Eqs.~(\ref{eq:c13}) and (\ref{eq:c33}), reduce to
\begin{align}
&C^{13}=\frac{1}{\sqrt{6}}\left( f_{I,II}^{11}(0,0,1,-1) - f_{II,I}^{11}(1,1,0,0) \right),\label{eq:c13_2qubit}\\
&C^{33}=\frac{1}{2} \Tr \left[R^{11}\right] + \frac{1}{2\sqrt{3}}\left( f_{I,II}^{11}(0,0,1,0) + f_{II,I}^{11}(1,0,0,0) \right).\label{eq:c33_2qubit}
\end{align}

\section{Calculating the depolarisation probability for qubit PBT with a maximally entangled resource}\label{secVI}

As a second example, we can calculate the channel enacted by PBT with a maximally entangled resource state. PBT with such a resource state enacts a depolarising channel \cite{ishizaka_quantum_2009}. Our analytical formulae for the components of the output Choi matrix give an easy way to calculate the depolarising probability of the channel simulated by $N$-port PBT. This probability is calculated in a similar way in Ref.~\cite{pirandola_fundamental_2019}, where it is referred to as the ``PBT number" for a given $N$, but without using the explicit formulae presented here.

The Choi matrix of a depolarising channel is
\begin{equation}
C_{\mathrm{dep}}=\begin{pmatrix}
\frac{1}{2}-\frac{\xi}{4} &0 &0 &\frac{1}{2}-\frac{\xi}{2}\\
0 &\frac{\xi}{4} &0 &0\\
0 &0 &\frac{\xi}{4} &0\\
\frac{1}{2}-\frac{\xi}{2} &0 &0 &\frac{1}{2}-\frac{\xi}{4}
\end{pmatrix},
\end{equation}
where $\xi$ is the depolarising probability of the channel. Since the channel has only one parameter, we only need to find one (non-zero) element of the Choi matrix in order to characterise it. We pick $C_{\mathrm{dep}}^{33}$ (the third element on the main diagonal); the expression for this component is given by Eq.~({\ref{eq:c33}}).

We start by finding $R^{11}$ for the maximally entangled resource $\left|\Phi^{\mathrm{Bell}}\right>\left<\Phi^{\mathrm{Bell}}\right|^{\otimes N}$, where $\left|\Phi^{\mathrm{Bell}}\right>=\frac{1}{\sqrt{2}}(\left|01\right>-\left|10\right>)$. We find
\begin{equation}
\begin{split}
R^{11}&=\langle 0|_{B_1} \mathrm{Tr}_{\bar{B_1}}[\left|\Phi^{\mathrm{Bell}}\right>\left<\Phi^{\mathrm{Bell}}\right|^{\otimes N}_{AB}] |0\rangle_{B_1}\\
&=\frac{1}{2^{N-1}}\langle 0|_{B_1}\left|\Phi^{\mathrm{Bell}}\right>\left<\Phi^{\mathrm{Bell}}\right|_{A_1 B_1}|0 \rangle_{B_1}\otimes \mathbb{I}_{\bar{A_1}}\\
&=\frac{1}{2^{N}}\left|1\right>\left<1\right|_{A_1}\otimes \left(\sum_{j,m,\alpha} \left|\Phi^{[N-1]}(j,m,\alpha)\right>\left<\Phi^{[N-1]}(j,m,\alpha)\right|_{\bar{A_1}} \right),
\end{split}
\end{equation}
where the sum is over all valid values of $j$, $m$, and $\alpha$. We can express $R^{11}$ in the $N$-spin basis using Eqs.~(\ref{eq:phi1}) and (\ref{eq:phi2}). This allows us to write the functions
\begin{align}
&f_{I,I}^{11}(s-\frac{1}{2},m-\frac{1}{2},s-\frac{1}{2},m-\frac{1}{2})=\frac{1}{2^N}\left[\Xi^{- +}(s,m-1)\right]^2=\frac{s-m+1}{2^N(2s+1)}\\
&\begin{aligned}
f_{I,II}^{11}(s-\frac{1}{2},m-\frac{1}{2},s+\frac{1}{2},m-\frac{1}{2})&=\frac{1}{2^N}\left[\Xi^{- +}(s,m-1)\Xi^{+ +}(s,m-1)\right]\\
&=-\frac{\sqrt{(s-m+1)(s+m)}}{2^N(2s+1)}
\end{aligned}\\
&f_{II,II}^{11}(s+\frac{1}{2},m-\frac{1}{2},s+\frac{1}{2},m-\frac{1}{2})=\frac{1}{2^N}\left[\Xi^{+ +}(s,m-1)\right]^2=\frac{s+m}{2^N(2s+1)},
\end{align}
noting also that $f_{I,II}^{11}=f_{II,I}^{11}$, since $R^{11}$ is a conditional density matrix (and therefore must be hermitian).

We can express the degeneracy for the $N-1$-spin basis as
\begin{equation}
\gamma(N-1,s)=\frac{(2s+1)(N-1)!}{\left(\frac{N-1}{2}-s\right)!\left(\frac{N+1}{2}+s\right)!}=\frac{2s+1}{N}\binom{N}{\frac{N-1}{2}-s},
\end{equation}
where the expression on the right hand side uses a binomial coefficient. We can therefore write
\begin{equation}
q_{+}^2 g_{I,I} +r_{-}^2 g_{II,II} +q_{+}r_{-} \left(g_{I,II}+g_{II,I}\right)=\frac{(s+m)(s-m+1)}{2^{N+1}N(2s+1)}\binom{N}{\frac{N-1}{2}-s}\left[(\lambda^{-}_{s-\frac{1}{2}})^{-\frac{1}{2}}-(\lambda^{+}_{s+\frac{1}{2}})^{-\frac{1}{2}}\right]^2,\label{eq: depol cont}
\end{equation}
where it is implicit that the indices for the $g$-functions are those found in the first sum in Eq.~(\ref{eq:c33}). We then carry out the sum
\begin{equation}
\sum_{m=-s}^s (s+m)(s-m+1)=\frac{2}{3}s(s+1)(2s+1).
\end{equation}
We expand the last term in Eq.~(\ref{eq: depol cont}), getting
\begin{equation}
\left[(\lambda^{-}_{s-\frac{1}{2}})^{-\frac{1}{2}}-(\lambda^{+}_{s+\frac{1}{2}})^{-\frac{1}{2}}\right]^2=8\frac{(N+2)-\sqrt{(N+2)^2-(2s+1)^2}}{(N+2)^2-(2s+1)^2}.
\end{equation}

We then calculate
\begin{equation}
\left(\frac{1}{2}+\frac{m}{N+1}\right)g_{II,II}=\frac{1}{2^N N(N+1)}\left(\frac{N-1}{2}+m\right)\left(\frac{N+1}{2}+m\right),
\end{equation}
where it is implicit that the indices for the $g$-function are those found in the second sum in Eq.~(\ref{eq:c33}). We perform the sum
\begin{equation}
\sum_{m=-\frac{N+1}{2}}^{\frac{N+1}{2}} \left(\frac{N-1}{2}+m\right)\left(\frac{N+1}{2}+m\right)=\frac{1}{3}N(N+1)(N+2).
\end{equation}

Substituting these expressions into Eq.~(\ref{eq:c33}), we get
\begin{equation}
C_{\mathrm{dep}}^{33}=\frac{1}{3\times2^{N-2}}\sum_{s=s_{\mathrm{min}}}^{\frac{N-1}{2}} \left[ s(s+1)\binom{N}{\frac{N-1}{2}-s}\frac{(N+2)-\sqrt{(N+2)^2-(2s+1)^2}}{(N+2)^2-(2s+1)^2} \right] + \frac{N+2}{3\times2^{N+1}},
\end{equation}
which immediately gives
\begin{equation}
\xi_N=\frac{1}{3\times2^{N-4}}\sum_{s=s_{\mathrm{min}}}^{\frac{N-1}{2}} \left[ s(s+1)\binom{N}{\frac{N-1}{2}-s}\frac{(N+2)-\sqrt{(N+2)^2-(2s+1)^2}}{(N+2)^2-(2s+1)^2} \right] + \frac{N+2}{3\times2^{N-1}},\label{eq:xi_expression}
\end{equation}
where $\xi_N$ is the depolarising probability of the $N$-port qubit PBT channel with a maximally entangled resource. We numerically observe that $\xi_N$ scales approximately with $\frac{1}{N}$ for large $N$.

\section{Simulating the amplitude damping channel}\label{secVII}
We know that in the limit of $N\to\infty$, a resource state comprised of $N$ copies of the Choi matrix of a given channel perfectly simulates that channel. This is because PBT over such a resource state is equivalent to passing the transmitted state through an identity channel followed by the desired channel. However, for finite $N$, it may be the case that there is a resource state that simulates a given channel better than $N$ copies of the Choi matrix or even perfectly. Our metric for judging which of two channels is a better simulation of a given channel is the diamond norm, $D_{\diamond}$, between the simulated channel and the channel simulating it. The diamond norm between channels $\mathcal{E}_1$ and $\mathcal{E}_2$ is defined by
\begin{align}
D_{\diamond}=\sup_{\phi} \Tr \left| \mathbb{I}\otimes\mathcal{E}_1(\phi)-\mathbb{I}\otimes\mathcal{E}_2(\phi) \right|,
\end{align}
where the supremum is taken over all input states $\phi$ (and where the identity is enacted on idler modes of $\phi$). Of particular interest are resource states with tensor-product structure (i.e. $N$ identical copies of a two-qubit state). The simple structure of such states makes it easier to carry out calculations on them for channel simulation. For instance, \cite{pirandola_fundamental_2017} found that the achievable secret key rate of a quantum channel can be upper bounded by the relative entropy of entanglement (REE) of a resource state that can be used to simulate that channel. If a state has tensor-product structure, the calculation of its REE can be simplified: the REE of such a state is $N$ times the REE of a single copy of the two-qubit state. Let us refer to all resource states with tensor-product structure as tensor-product resources.

One channel of interest is the amplitude damping (AD) channel. This channel is characterised by the Choi matrix (for the input state $\left|\Phi^{\mathrm{Bell}}\right>=\frac{1}{\sqrt{2}}(\left|01\right>-\left|10\right>)$)
\begin{align}
R(p)=\begin{pmatrix}
\frac{p}{2} &0 &0 &0\\
0 &\frac{1-p}{2} &-\frac{\sqrt{1-p}}{2} &0\\
0 &-\frac{\sqrt{1-p}}{2} &\frac{1}{2} &0\\
0 &0 &0 &0
\end{pmatrix},\label{eq:choi_res}
\end{align}
where $p$ is the probability of a qubit with value one being flipped to a zero. One possible type of resource state is comprised of $N$ copies of this state, $R(p_1)^{\otimes N}$, where $p_1$ is the damping probability of the AD channel used to generate the resource state, i.e. the resource state is $N$ copies of the output Choi matrix of an AD channel with damping probability $p_1$. Note that this is not the same Bell state that we have been using to define the Choi matrix previously; we have previously used the input state $\frac{1}{\sqrt{2}}(\left|00\right>+\left|11\right>)$. We have chosen the state $\left|\Phi^{\mathrm{Bell}}\right>$ in this case because it is the resource state $\left|\Phi^{\mathrm{Bell}}\right>\left<\Phi^{\mathrm{Bell}}\right|^{\otimes N}$ that simulates the identity channel (due to the structure of the measurement). Consequently, it is the resource $R(p_1)^{\otimes N}$ that asymptotically gives a perfect simulation of the AD channel. Note that a different choice of Bell state would result in the simulated channel being a Pauli unitary and, similarly, the Choi matrices of a channel defined by different input Bell states are equivalent up to a Pauli unitary on one of the modes. Let $p_0$ be the damping probability of the AD channel that we are trying to simulate; this need not necessarily be equal to $p_1$. We denote the Choi matrix of the PBT channel with resource state $\phi$, $PBT[\phi]$. Applying the explicit expressions that we have derived, we find
\begin{align}
PBT[R(p_1)^{\otimes N}]=\begin{pmatrix}
\frac{1}{2}-\frac{\xi_N}{4}(1-p_1) &0 &0 &\left(\frac{1}{2}-\frac{\xi_N}{2}\right)\sqrt{1-p_1}\\
0 &\frac{\xi_N}{4}(1-p_1) &0 &0\\
0 &0 &p_1\left(\frac{1}{2}-\frac{\xi_N}{4}\right)+\frac{\xi_N}{4} &0\\
\left(\frac{1}{2}-\frac{\xi_N}{2}\right)\sqrt{1-p_1} &0 &0 &(1-p_1)\left(\frac{1}{2}-\frac{\xi_N}{4}\right)
\end{pmatrix}. \label{eq:choi of choi_res}
\end{align}
We will refer to such a resource state ($N$ copies of the Choi matrix of an AD channel, with damping probability generally different from that of the simulated channel) as a Choi resource.

Consider the special case of $p_1=p_0$ (simulating an AD channel with $N$ copies of its own output Choi matrix); it has been shown that in this case, the diamond norm of the simulated channel from the simulating channel is the same as the trace norm between the Choi matrices \cite{pirandola_fundamental_2019}. We will denote the diamond norm using this resource as $D_{\diamond}^0$; it is given by
\begin{align}
D_{\diamond}^0=\xi_N \left( \frac{1-p_0}{2} +\sqrt{1-p_0} \right),
\end{align}
where $\xi_N$ is the depolarisation probability when carrying out PBT with a maximally entangled resource state, as given by Eq.~(\ref{eq:xi_expression}). $\xi_N\leq \frac{6-\sqrt{3}}{6}\simeq0.71$, since this is the value for 2 ports. $D_{\diamond}^0$ provides a useful benchmark, since we know it converges to 0 in the limit of infinite ports, and hence $R(p_0)^{\otimes N}$ is a common choice of resource state for calculations involving channel simulation. For instance, in \cite{pirandola_fundamental_2019}, resource states composed of $N$ copies of the Choi matrix of the simulated channel were used to obtain a general bound on channel discrimination, and this bound was specifically applied to the AD channel.

In the asymptotic limit, in the case of $p_1=p_0$, the output Choi matrix in Eq.~(\ref{eq:choi of choi_res}) tends to the Choi matrix of the simulated channel, as expected. However, for finite $N$, a lower $D_{\diamond}$ can be achieved by choosing a value of $p_1$ for the resource state different from $p_0$ (the damping probability of the channel we are simulating).

Let us consider for which values of $p_1$ we can know the diamond norm exactly. We have upper and lower bounds on the diamond norm between (qubit) channels with Choi matrices $X$ and $Y$ given by \cite{nechita_almost_2018}:
\begin{align}
\Tr \left|X-Y\right| \leq D_{\diamond} \leq 2\left\lVert \Tr_2\left|X-Y\right|\right\rVert_\infty,
\end{align}
where the trace is taken over the mode which passed through the channel. These two bounds are equal (and therefore give the exact diamond norm) if the matrix $\Tr_2\left|X-Y\right|$ is scalar (proportional to the identity matrix). The difference between the Choi matrices of the simulated and simulating channels, in this case, is
\begin{align}
&PBT[R(p_1)^{\otimes N}]-R'(p_0)=\begin{pmatrix}
-e_1 &0 &0 &-c\\
0 &e_1 &0 &0\\
0 &0 &e_2 &0\\
-c &0 &0 &-e_2
\end{pmatrix},\label{eq:dif mat}\\
&e_1=\frac{\xi_N}{4}(1-p_1),\label{eq:e1}\\
&e_2=e_1-\frac{p_0-p_1}{2},\label{eq:e2}\\
&c=\frac{1}{2}\left(\sqrt{1-p_0}-(1-\xi_N)\sqrt{1-p_1}\right),
\end{align}
where $R'$ is the Choi matrix for the input state $\frac{1}{\sqrt{2}}(|00\rangle+|11\rangle)$.
If $e_1=\pm e_2$, the modulus of the matrix, with the trace taken over the second mode, will be scalar. This is true in two cases:
\begin{align}
&p_1=p_0,\label{eq:p exp 1}\\
&p_1=\frac{p_0-\xi_N}{1-\xi_N}.\label{eq:p exp 2}
\end{align}

\begin{figure}[ptb]
\centering
\includegraphics[width=1\linewidth]{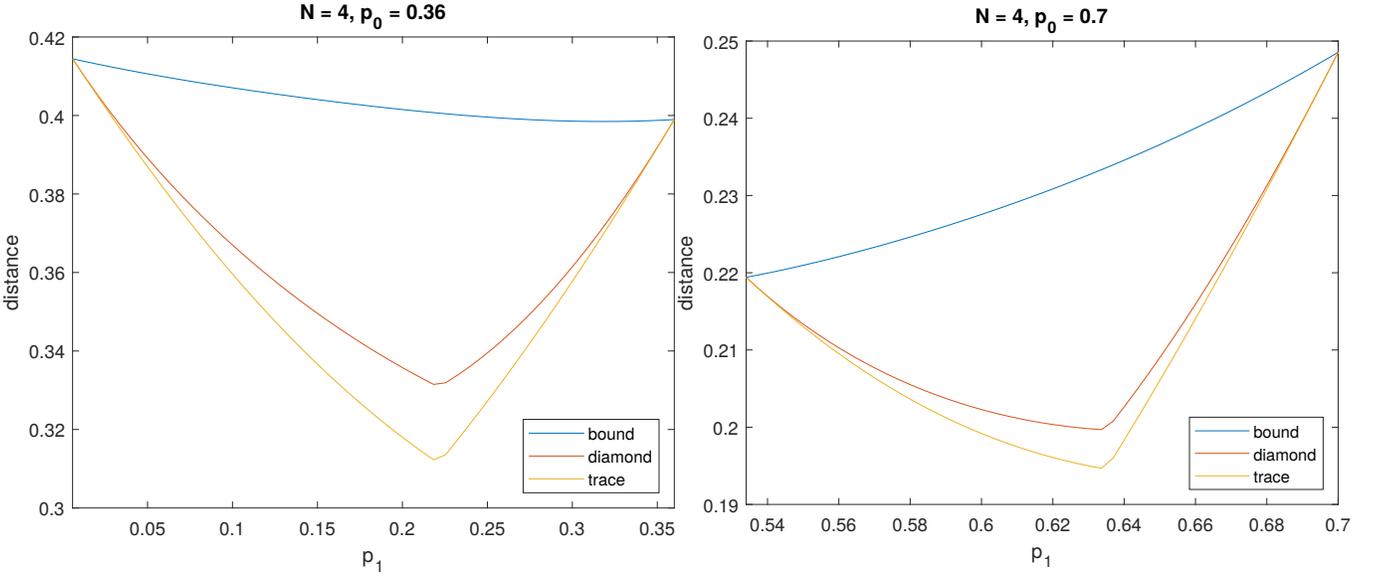}\caption{The trace norm, the numerically found diamond norm and the analytical upper bound on the diamond norm from \cite{nechita_almost_2018} are plotted against $p_1$, the damping value of the AD channel used to produce the resource state, for the resource given in Eq.~(\ref{eq:choi_res}). The plot with $p_0=0.36$ lies in the regime where $p_1=p_0$ gives a better simulation than $p_1=\frac{p_0-\xi_N}{1-\xi_N}$, and the plot with $p_0=0.7$ lies in the regime where the opposite is true. In both cases, the actual minimum of the diamond norm lies between these points, and lies near the minimum of the trace norm. In both cases, this minimum of the trace norm lies at exactly $p_1=\frac{2p_0-\xi_N}{2-\xi_N}$.}
\label{fig:high low choi_res}
\end{figure}

The first case is the known case of $N$ copies of the Choi matrix of the simulated channel. In the second case, we find that the diamond norm, $D_{\diamond}^1$, is given by
\begin{align}
D_{\diamond}^1=\frac{1}{2}\left(\frac{(1-p_0)\xi_N}{1-\xi_N}+\sqrt{4(1-p_0)\left(1-\sqrt{1-\xi_N}\right)^2+\frac{(1-p_0)^2\xi_N^2}{(1-\xi_N)^2}}\right).\label{eq:choi_diamond}
\end{align}
For sufficiently low values of $\xi_N$ and sufficiently high values of $p_0$, this second expression for the diamond norm, $D_{\diamond}^1$ is lower than $D_{\diamond}^0$. Specifically, we find that there is a function in $\xi_N$ separating the two regimes. This function crosses $p_0=0$ at a $\xi_N$ value of about 0.237, and for values of $\xi_N<0.237$, the second expression is always lower (except in the trivial case of $p_0=1$). $\xi_N<0.237$ for a number of ports equal to or greater than 6, so for $N\geq6$, $D_{\diamond}^1\leq D_{\diamond}^0$. Note that if $p_0<\xi_N$, this second point does not exist, since that would require a negative value of $p_1$. The plots in Fig.~\ref{fig:high low choi_res} illustrate these two regimes in the case of 4 ports. We therefore have a resource that simulates a given AD channel better than $N$ copies of the Choi matrix of that channel, for any finite number of ports, with an analytical expression for the diamond norm between the channels.

Asymptotically (in $N$), the right hand side of Eq.~(\ref{eq:p exp 2}) tends to the right hand side of Eq~\ref{eq:p exp 1}, since $\xi_N$ tends to 0. This is as expected, since we know that the Choi resource with $p_1=p_0$ simulates the AD channel perfectly in the asymptotic limit of $N$.

Although we have two points for which the diamond norm is known exactly, this does not mean that the minimum diamond norm for simulating a given channel lies at either of these two points. In fact, we find numerically that the minimum of the diamond norm often lies near the minimum of the trace norm between the Choi matrices, rather than at either of these known points. We also find (in an appendix) that for all $p_0\leq v_1$, where $v_1$ is a function of $\xi_N$ that is always greater than $\frac{2}{5}$, the minimum of the trace norm lies at $\frac{2p_0-\xi_N}{2-\xi_N}$, and that for all $p_0\leq v_2$, where $v_2$ is a function of $\xi_N$ that is always greater than $\frac{2}{3}$, the minimum of the trace norm lies between $p_1=\frac{p_0-\xi_N}{1-\xi_N}$ and $p_1=\frac{2p_0-\xi_N}{2-\xi_N}$.

\begin{figure}[ptb]
\centering
\includegraphics[width=1\linewidth]{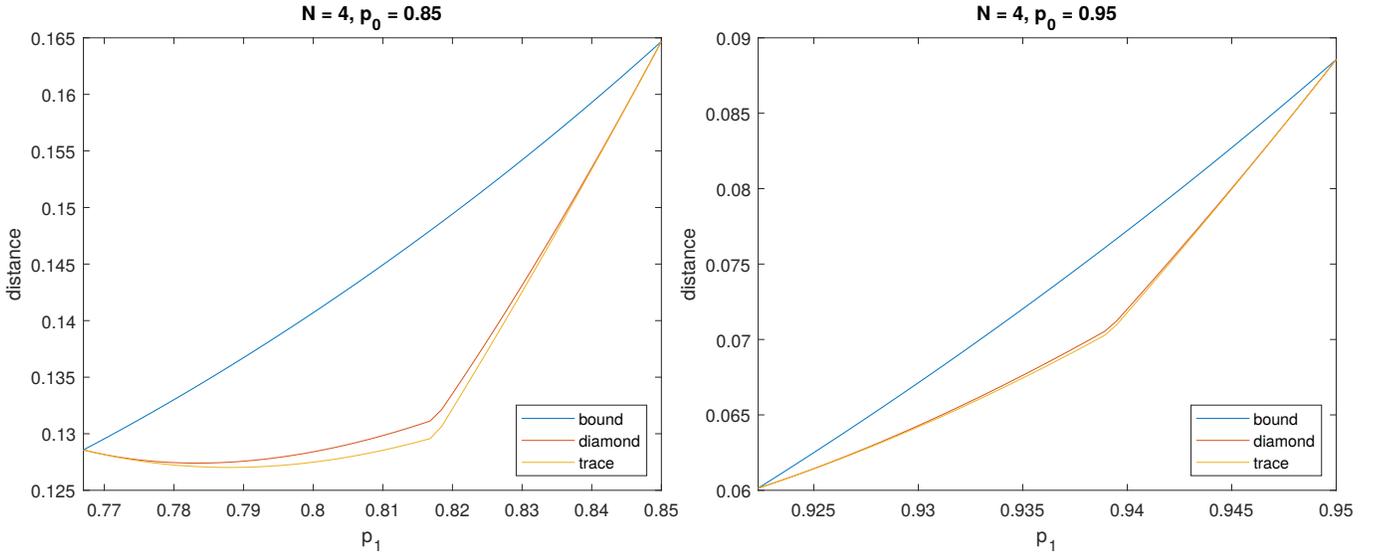}\caption{The trace norm, the numerically found diamond norm and the analytical upper bound on the diamond norm from \cite{nechita_almost_2018} are plotted against $p_1$, the damping value of the AD channel used to produce the resource state, for the resource given in Eq.~(\ref{eq:choi_res}). In both of the cases shown, the minimum of the trace norm no longer lies at $p_1=\frac{2p_0-\xi_N}{2-\xi_N}$, but rather at a lower value of $p_1$. In the case of $p_0=0.85$, the minimum of the trace norm (and therefore of the diamond norm) still lies between the two points for which the diamond norm is exactly known ($p_1=\frac{p_0-\xi_N}{1-\xi_N}$ and $p_1=p_0$), whereas for $p_0=0.95$, this is no longer the case.}
\label{fig:past bounds choi_res}
\end{figure}

If the minimum of the trace norm lies between $p_1=\frac{p_0-\xi_N}{1-\xi_N}$ and $p_1=p_0$, the two points at which the diamond norm is equal to the trace norm, we are guaranteed that the minimum of the diamond norm will fall between those two points, since the trace norm, which lower bounds the diamond norm, will have no local minima outside of these points. This means that the trace norm will have a negative gradient at every point below $p_1=\frac{p_0-\xi_N}{1-\xi_N}$ and a positive gradient at every point above $p_1=p_0$. The plots in Fig.~\ref{fig:past bounds choi_res} show values of $p_0$ for which the minimum of the trace norm does not lie at $p_1=\frac{2p_0-\xi_N}{2-\xi_N}$.

Whilst the Choi resource with $p_1$ chosen to minimise the diamond norm simulates the AD channel better than the case of $p_1=p_0$, the two resources tend towards each other as $N$ increases. A resource state of interest would be one that has tensor-product structure, simulates some AD channel better than the Choi resource and is distinct from the Choi resource for all $p_1$ values. We find that such a resource exists. Let $R_{\mathrm{new}}(a)$ be a two-qubit state, defined by
\begin{align}
R_{\mathrm{new}}(a)=\begin{pmatrix}
0 &0 &0 &0\\
0 &a &-\sqrt{a(1-a)} &0\\
0 &-\sqrt{a(1-a)} &1-a &0\\
0 &0 &0 &0
\end{pmatrix}\label{eq:new_res},
\end{align}
where $a$ is a parameter characterising the density matrix. Consider the resource state $R_{\mathrm{new}}(a)^{\otimes N}$ ($N$ copies of $R_{\mathrm{new}}(a)$, such that each port is a copy of $R_{\mathrm{new}}(a)$). This is a tensor-product resource and the state of each port is clearly different from the state in Eq.~(\ref{eq:choi_res}) for all parameter values except for the case of $p=0$ and $a=\frac{1}{2}$. This resource state illustrates the importance of the explicit expressions for the components of the Choi matrix resulting from PBT: whilst it would be possible to calculate $PBT[R(p)^{\otimes N}]$ by applying an AD channel to the (known) output of the PBT channel using a maximally entangled resource, the same technique cannot be used to calculate $PBT\left[R_{\mathrm{new}}(a)^{\otimes N}\right]$.

Carrying out PBT using this resource state, which we will call the alternate resource, results in the Choi matrix:
\begin{align}
&\begin{aligned}
PBT\left[R_{\mathrm{new}}(a)^{\otimes N}\right]=\begin{pmatrix}
x &0 &0 &z\\
0 &\frac{1}{2}-x &0 &0\\
0 &0 &y &0\\
z &0 &0 &\frac{1}{2}-y
\end{pmatrix},\label{eq:new_choi}
\end{aligned}\\
&\begin{aligned}
x=\sum_{s=s_{\mathrm{min}}}^{\frac{N-1}{2}}\sum_{m=-s}^{s}a^{\frac{N+1}{2}+m}(1-a)^{\frac{N-1}{2}-m}\frac{N!\left[\left(\frac{N+1}{2}-s\right)^{-\frac{1}{2}}(s-m)+\left(\frac{N+3}{2}+s\right)^{-\frac{1}{2}}(s+m+1)\right]^2}{2\left(\frac{N-1}{2}-s\right)!\left(\frac{N+1}{2}+s\right)!(2s+1)}\\
+\sum_{m=-\frac{N+1}{2}}^{\frac{N+1}{2}}a^{\frac{N+1}{2}+m}(1-a)^{\frac{N-1}{2}-m}\frac{\left(\frac{N+1}{2}+m\right)\left(\frac{N+1}{2}-m\right)}{2N(N+1)},
\end{aligned}\\
&\begin{aligned}
y=\sum_{s=s_{\mathrm{min}}}^{\frac{N-1}{2}}\sum_{m=-s}^{s}a^{\frac{N-1}{2}+m}(1-a)^{\frac{N+1}{2}-m}\frac{N!(s+m)(s-m+1)\left[\left(\frac{N+1}{2}-s\right)^{-\frac{1}{2}}-\left(\frac{N+3}{2}+s\right)^{-\frac{1}{2}}\right]^2}{2\left(\frac{N-1}{2}-s\right)!\left(\frac{N+1}{2}+s\right)!(2s+1)}\\
+\sum_{m=-\frac{N+1}{2}}^{\frac{N+1}{2}}a^{\frac{N-1}{2}+m}(1-a)^{\frac{N+1}{2}-m}\frac{\left(\frac{N-1}{2}+m\right)\left(\frac{N+1}{2}+m\right)}{2N(N+1)},
\end{aligned}\\
&\begin{aligned}
z=\sum_{s=s_{\mathrm{min}}}^{\frac{N-1}{2}}\sum_{m=-s}^{s}&\frac{a^{\frac{N}{2}+m}(1-a)^{\frac{N}{2}-m}N!}{2\left(\frac{N-1}{2}-s\right)!\left(\frac{N+1}{2}+s\right)!(2s+1)}\left[\left(\frac{N+1}{2}-s\right)^{-1}(s^2-m^2)\right.\\
&\left.+2\left(\frac{N+1}{2}-s\right)^{-\frac{1}{2}}\left(\frac{N+3}{2}+s\right)^{-\frac{1}{2}}(s^2+m^2+s)+\left(\frac{N+3}{2}+s\right)^{-1}((s+1)^2-m^2)\right]\\
&-\sum_{m=-\frac{N+1}{2}}^{\frac{N+1}{2}}a^{\frac{N}{2}+m}(1-a)^{\frac{N}{2}-m}\frac{\left(\frac{N+1}{2}+m\right)\left(\frac{N+1}{2}-m\right)}{2N(N+1)},
\end{aligned}
\end{align}
where $s_{\mathrm{min}}$ is 0 for odd $N$ and $\frac{1}{2}$ for even $N$. The elements of the Choi matrix have been calculated using the expressions in Eqs.~(\ref{eq:c11}) to (\ref{eq:c33}). We can therefore write
\begin{align}
PBT\left[R_{\mathrm{new}}(a)^{\otimes N}\right]-R'(p_0)=\begin{pmatrix}
x-\frac{1}{2} &0 &0 &z-\frac{\sqrt{1-p_0}}{2}\\
0 &\frac{1}{2}-x &0 &0\\
0 &0 &y-\frac{p_0}{2} &0\\
z-\frac{\sqrt{1-p_0}}{2} &0 &0 &\frac{p_0}{2}-y
\end{pmatrix},
\end{align}

\begin{figure}[ptb]
\centering
\includegraphics[width=1\linewidth]{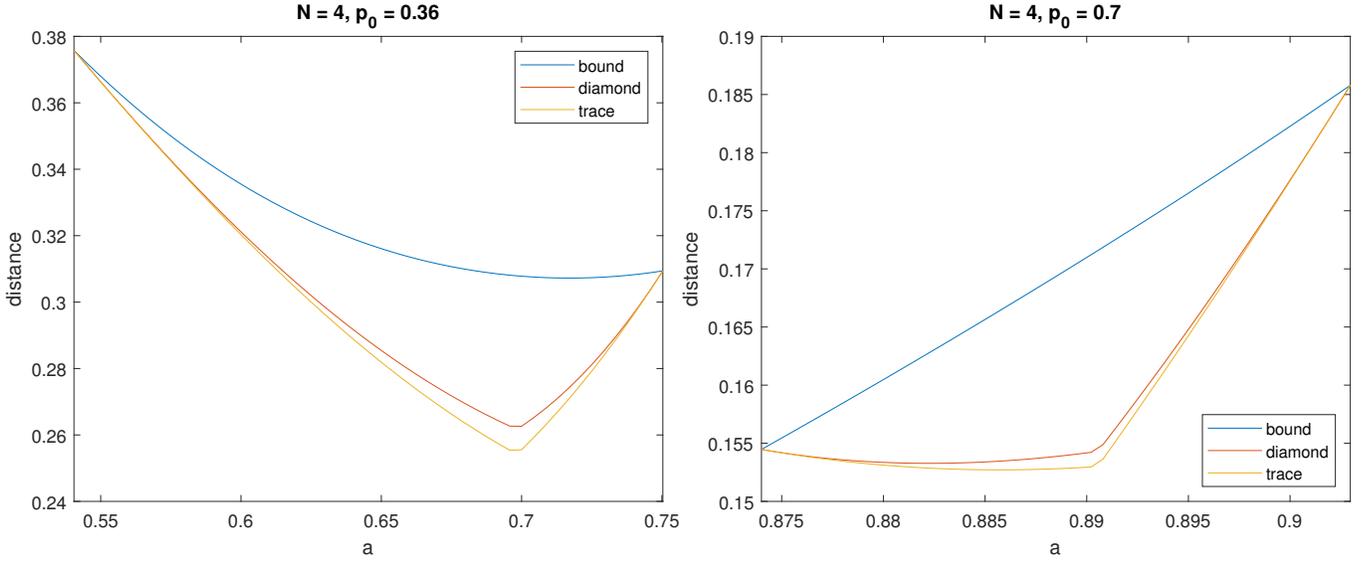}\caption{The trace norm, the numerically found diamond norm and the analytical upper bound on the diamond norm from \cite{nechita_almost_2018} are plotted against $a$, the parameter that parametrises the state in Eq.~(\ref{eq:new_res}). Comparing with Fig.~\ref{fig:high low choi_res}, we can see that at the ``known points" where the diamond norm is known analytically (where the trace norm coincides with the diamond norm), the diamond norm is significantly lower for the resource $R_{\mathrm{new}}(a)^{\otimes N}$ than at the known points for the Choi resource. Further, the minimum diamond norm for this new resource is significantly lower than the minimum diamond norm for the Choi resource.}
\label{fig:high low new_res}
\end{figure}

Again, we can find the values of $a$ at which this matrix is scalar by finding the points at which $x-\frac{1}{2}=\pm \left(y-\frac{p_0}{2}\right)$. In this case, however, we have a more complicated expression in terms of $a$ and $p_0$, which depends on $N$, making it difficult to find a general (for arbitrary $N$) expression for the diamond norm at these points where the diamond norm is known exactly (however it is simple to find the expression for fixed $N$).

Using this resource, we can prove that for all $N$ and for some range of $p_0$ values, there exists some tensor-product resource, which is distinct from $R(p)^{\otimes N}$, for which the diamond norm from the AD channel can be found analytically and is smaller than the diamond norm using the resource state $R(p)^{\otimes N}$ for both $p=p_0$ and $p=\frac{p_0-\xi_N}{1-\xi_N}$. This means that, for any finite value of $N$, there are some (low) values of $p_0$ for which we can find a tensor-product resource state that gives a diamond norm from the AD channel lower than either $D_{\diamond}^0$ or $D_{\diamond}^1$. This is demonstrated in Fig.~\ref{fig:high low new_res}, for $N=4$, using the resource state $R_{\mathrm{new}}(a)^{\otimes N}$, and is proven in an appendix.

\begin{figure}[ptb]
\centering
\includegraphics[width=1\linewidth]{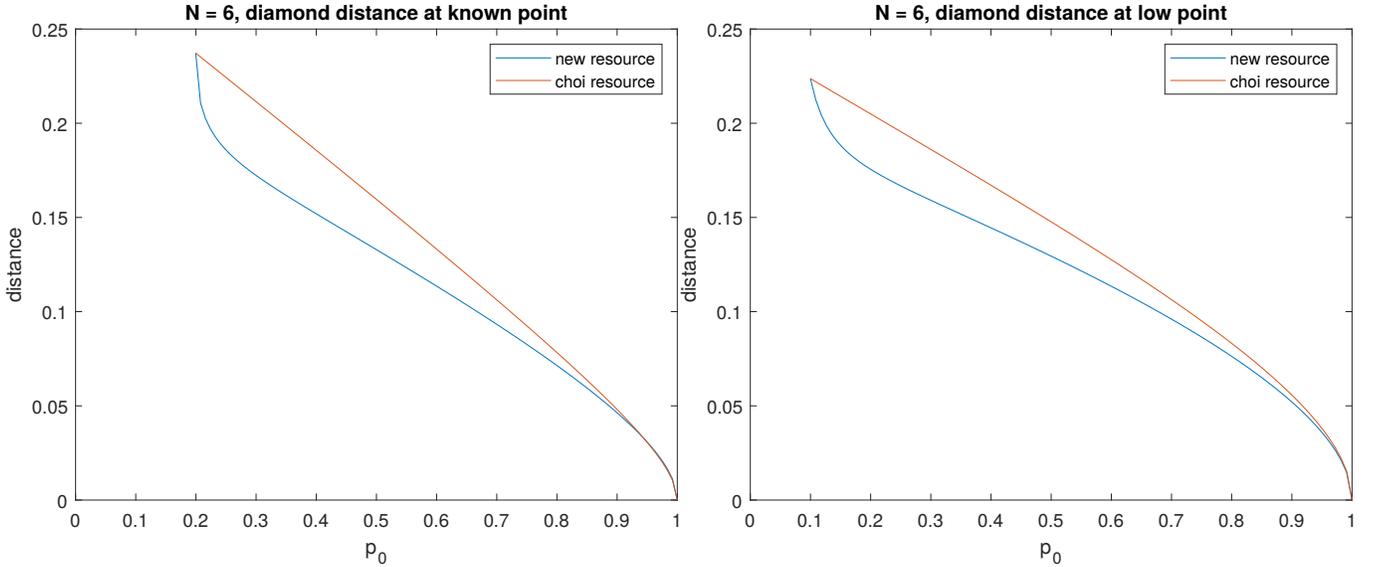}\caption{The diamond norm is plotted against the damping probability of the AD channel being simulated for PBT with the resource state $R_{\mathrm{new}}(a)^{\otimes N}$ (new resource) and the resource state $R(p_1)^{\otimes N}$ (Choi resource). In the left-hand plot, we choose $p_1=\frac{p_0-\xi_N}{1-\xi_N}$ and choose $a$ such that $x(a)-y(a)=\frac{1-p_0}{2}$, so that the trace norm coincides with the diamond norm. In the right hand plot, we choose $p_1=\frac{2p_0-\xi_N}{2-\xi_N}$ and choose $a$ such that $y(a)=\frac{p_0}{2}$; these are close to the optimal parameters to minimise the diamond norm. In both cases, we start at the minimum value of $p_0$ for which $p_1$ is non-negative. The new resource is better than the Choi resource for a large range of $p_0$ values, and especially for low $p_0$.}
\label{fig:resource comparison}
\end{figure}

For low $N$, the alternate resource beats the Choi resource over a large range of $p_0$ values, and by a significant amount. This can be seen for the case of $N=6$ in Fig.~\ref{fig:resource comparison}. Note that at $a=\frac{1}{2}$ and $p=0$, the two resources are the same, and these parameter values are the starting points of the graphs in the figure.

Similarly to the case of the Choi resource, we find numerically that for a large range of $p_0$ values, the value of $a$ that gives the minimum of the trace norm coincides with the value that minimises the diamond norm, and is the $a$ value for which $y-\frac{p_0}{2}=0$ (just as, for the Choi resource, the minimum of the trace norm occurs at the value of $p$ that sets $e_2=0$, for all $p_0<\frac{2}{5}$). Numerically we find a trend that there exists a range of $p_0$ values such that the resource state $R_{\mathrm{new}}(a)^{\otimes N}$, with $a$ chosen so that $y=\frac{p_0}{2}$, gives a better simulation of the AD channel (lower diamond norm) than $R(p_1)^{\otimes N}$, for any value of $p_1$. However, this range of $p_0$ values becomes increasingly small as $N$ increases. This has been numerically confirmed for $N<11$. Specifically, this occurs for low $p_0$.

The explicit expressions for the Choi matrix of the PBT channel therefore allow us to calculate the diamond norm for a resource that simulates certain AD channels better than a tensor-product of Choi matrices.

\section{Conclusion}\label{secVIII}
Qubit PBT simulates a quantum channel on the teleported qubit, with the channel depending on the resource state used. Using Eqs.~(\ref{eq:c11}) to (\ref{eq:total_choi}), we can find the Choi matrix for the channel simulated by a given resource state. We assume this resource state to be symmetric under exchange of labels, since this assumption does not restrict the simulable channels. We also provide a simple algorithm for converting to the alternative channel representation of Kraus operators. We show how the Choi matrix can be easily calculated in the two port case, giving simplified expressions (namely, Eqs.~(\ref{eq:c11_2qubit}), (\ref{eq:c13_2qubit}) and (\ref{eq:c33_2qubit})).

In Eqs.~(\ref{eq:kraus1}) and (\ref{eq:kraus2}), we give the Kraus operators that describe the PBT protocol itself (for a fixed number of ports, the square-root measurement and a resource state that is symmetric under exchange of labels). These Kraus operators characterise the map from the $2N$-qubit resource state to the two-qubit Choi matrix, and thus offer a complete description of the PBT protocol. This is a complete analytical characterisation that could be efficiently exploited in Ref.~\cite{banchi_convex_2020} where techniques of machine learning and semi-definite programming are employed to find the optimal resource state for PBT (and other teleportation protocols).

We consider simulating the amplitude damping channel with PBT and find that, for finite numbers of ports, using $N$ copies of the Choi matrix of the simulated channel as the resource state gives a higher diamond norm than using $N$ copies of the Choi matrix of a different AD channel. We also find that there exist resource states with tensor-product structure that simulate the AD channel better than any Choi resource, in the low damping range.

In this paper, we only present results for the qubit case. Future work could explore PBT in the qudit or continuous variable cases. In the qudit case, this is complicated by the Clebsch-Gordan coefficients, which do not take the simple form they take in the qubit case. Clarifying the mathematical aspects of PBT is important for the fundamental role that this protocol plays in various areas of quantum information theory, not only in problems of ultimate channel discrimination~\cite{pirandola_fundamental_2019} but also in communication problems such as position-based quantum cryptography~\cite{beigi_simplified_2011,buhrman_position-based_2011}.

\smallskip
\textbf{Acknowledgments.}~This work was funded by the EPSRC Quantum Communications Hub (Grants No. EP/M013472/1 and No. EP/T001011/1) and the European Union via ``Continuous Variable Quantum Communications'' (CiViQ, Grant agreement No 820466) and ``Quantum readout techniques and technologies'' (QUARTET, Grant agreement No 862644). L.B. acknowledges support by the program ``Rita Levi Montalcini'' for young researchers.

\bibliography{bibfile}

\appendix
\section{Appendices}

\subsection{Proof of the location of the minima of the trace norm, for the Choi resource}
Let us calculate the trace norm by finding the eigenvalues of the matrix resulting from taking the difference of the Choi matrices of the simulated and simulating channel (i.e. the right hand side of Eq.~(\ref{eq:dif mat})). This matrix has eigenvalues $e_i$, where $e_1$ and $e_2$ have already been given in Eqs.~(\ref{eq:e1}) and (\ref{eq:e2}). The remaining eigenvalues are:
\begin{align}
&e_3=-\frac{1}{2}\left((e_1+e_2)+\sqrt{(e_1-e_2)^2+4c^2}\right),\\
&e_4=-\frac{1}{2}\left((e_1+e_2)-\sqrt{(e_1-e_2)^2+4c^2}\right).
\end{align}
The trace norm is the sum of the absolute values of the eigenvalues. We can show that $e_3$ is always negative and $e_4$ is always positive. We start by showing that $|e_1+e_2|\leq\sqrt{(e_1-e_2)^2+4c^2}$. Note that $e_1$ is a linear function of $p_1$ that is always positive and that $e_2$ is a linear function of $p_1$ that goes to 0 at $p_1=\frac{2p_0-\xi_N}{2-\xi_N}$, and is negative for $p_1$ less than this value. For $p_1=\frac{p_0-\xi_N}{1-\xi_N}$, $e_1+e_2=0$, and above this value of $p_1$, it is positive. We can therefore show that $2c\geq|e_1+e_2|$ in the regime in which $e_1+e_2$ is positive, using
\begin{align}
&\frac{d(2c)}{dp_1}=\frac{1-\xi_N}{2\sqrt{1-p_1}},\frac{d(e_1+e_2)}{dp_1}=\frac{1-\xi_N}{2},\frac{d(2c)}{dp_1}\geq\frac{d(e_1+e_2)}{dp_1},\\
&2c|_{p_1=\frac{p_0-\xi_N}{1-\xi_N}}=\sqrt{1-p_0}-(1-\xi_N)\sqrt{1-\frac{p-\xi_N}{1-\xi_N}}=\sqrt{1-p_0}(1-\sqrt{1-\xi_N})\geq 0.
\end{align}
Since the gradient of $2c$ is always larger than the gradient of $e_1+e_2$ in this regime, and since $c$ is positive at $p_1=\frac{p_0-\xi_N}{1-\xi_N}$, whilst $e_1+e_2$ is equal to 0, $2c\geq|e_1+e_2|$ for $p_1\geq\frac{p_0-\xi_N}{1-\xi_N}$. For $p_1<\frac{p_0-\xi_N}{1-\xi_N}$, $e_1-e_2=\frac{p_0-p_1}{2}\geq|e_1+e_2|$, because $e_2$ is negative in this region. Hence, at all points,
\begin{align}
|e_1+e_2|\leq\textrm{max}[e_1-e_2,2c]\leq\sqrt{(e_1-e_2)^2+4c^2}.
\end{align}
As a result, $e_3$ is always negative and $e_4$ is always positive. We therefore find
\begin{align}
|e_3|+|e_4|=\sqrt{(e_1-e_2)^2+4c^2}.
\end{align}
$|e_1|+|e_2|$ has two regimes, corresponding to $p_1\leq\frac{2p_0-\xi_N}{2-\xi_N}$ and $p_1>\frac{2p_0-\xi_N}{2-\xi_N}$. In the first regime, $|e_1|+|e_2|=\frac{p_0-p_1}{2}$, and in the second, $|e_1|+|e_2|=\frac{\xi_N}{2}(1-p_1)-\frac{p_0-p_1}{2}$. The gradient of $|e_1|+|e_2|$ is $-\frac{1}{2}$ in the first regime and $\frac{1-\xi_N}{2}$ in the second regime, with a discontinuity at $p_1=\frac{2p_0-\xi_N}{2-\xi_N}$. Taking the second derivative of $(e_1-e_2)^2+4c^2$, we find that it is always positive, so the gradient of $|e_3|+|e_4|$ is always increasing, and hence $|e_3|+|e_4|$ has at most one minimum.

The gradient of $|e_3|+|e_4|$ is given by
\begin{align}
\frac{d|e_3|+|e_4|}{dp_1}=\frac{p_1-p_0+2(1-\xi_N)\left(\sqrt{\frac{1-p_0}{1-p_1}}-(1-\xi_N)\right)}{4\sqrt{\frac{p_0-p_1}{2}^2+(\sqrt{1-p_0}-(1-\xi_N)\sqrt{1-p_1})^2}},
\end{align}
and the gradient of the total trace norm, $D_{trace}$, is given by
\begin{align}
&\left.\frac{d D_{trace}}{d p_1}\right|_{p_1<\frac{2p_0-\xi_N}{2-\xi_N}}=\frac{d|e_3|+|e_4|}{dp_1}-\frac{1}{2},\\
&\left.\frac{d D_{trace}}{d p_1}\right|_{p_1>\frac{2p_0-\xi_N}{2-\xi_N}}=\frac{d|e_3|+|e_4|}{dp_1}+\frac{1-\xi_N}{2}.
\end{align}
Note that the expressions for the gradient of the trace norm are different in each regime (on either side of the discontinuity).

Consider the case in which the minimum of $|e_3|+|e_4|$ occurs ``after" the discontinuity (i.e. at $p_1>\frac{2p_0-\xi_N}{2-\xi_N}$). There are two possibilities: if the (second) expression for the gradient of the trace norm assessed at $p_1=\frac{2p_0-\xi_N}{2-\xi_N}$ is negative, the minimum of the trace norm will lie in the region $p_1>\frac{2p_0-\xi_N}{2-\xi_N}$, whereas if it is positive, there is no stationary point and the minimum of the trace norm is located exactly at the discontinuity. By numerically minimising the expression for the gradient assessed at the discontinuity over $p$ (between 0 and 1) and over $\xi_N$ (between 0 and $\frac{6-\sqrt{3}}{6}$), we find that it is always positive. Hence, if the minimum of $|e_3+e_4|$ occurs at $p_1>\frac{2p_0-\xi_N}{2-\xi_N}$, the minimum of the trace norm lies at $\frac{2p_0-\xi_N}{2-\xi_N}$. Note that this is the point at which $e_2=0$.

Similarly, if the minimum of $|e_3|+|e_4|$ occurs ``before" the discontinuity, but the (first) expression for the gradient of the trace norm remains negative up to the discontinuity, the minimum of the trace norm will be at the discontinuity. Solving for this gradient to equal 0, we get a function in $\xi_N$ and $p_0$, giving the value of $p_1$ at which the minimum of the trace norm occurs (or would occur, if it is after the discontinuity). When this value becomes less than $\frac{2p_0-\xi_N}{2-\xi_N}$, the minimum of the trace norm lies at the value of this function, rather than at the discontinuity. We can find the value of $p_0$ at which this occurs for a given value of $\xi_N$. This is a function of $\xi_N$ only. Higher values of $\xi_N$ require higher values of $p_0$, and the minimum value of $p_0$ for which the minimum of the trace norm can occur in the the region $p_1<\frac{2p_0-\xi_N}{2-\xi_N}$ is $\frac{2}{5}$. For all $p_0<\frac{2}{5}$, the minimum trace norm always lies at $p_1=\frac{2p_0-\xi_N}{2-\xi_N}$.

We can find the value of $p_0$ at which the minimum of the trace norm crosses the line $p_1=\frac{p_0-\xi_N}{1-\xi_N}$, which we denote $p_0^{cross}$. We find that we have another function of $\xi_N$:
\begin{align}
p_0^{cross}=\frac{1+4\xi_N-8\xi_N^2+5\xi_N^3+(1-\xi_N)^{\frac{7}{2}}-\xi_N^4}{3-3\xi_N+\xi_N^2}.
\end{align}
This function has a minimum value of $\frac{2}{3}$, at $\xi_N=0$. Note that if $p_0\leq p_0^{cross}$, the gradient of $|e_3|+|e_4|$ is always negative in the range $p_1<\frac{p_0-\xi_N}{1-\xi_N}$ and is always positive in the range $p_1>p_0$, and hence the same is true of the gradient of the trace norm. Hence, for all $p_0\leq\frac{2}{3}$, we are guaranteed that the minimum of the diamond norm lies between $p_1=\frac{p_0-\xi_N}{1-\xi_N}$ and $p_1=\frac{2p_0-\xi_N}{2-\xi_N}$. For more detail, see the supplementary information \cite{zenodo}.

\subsection{Proof the alternate resource simulates the amplitude damping channel better than the Choi resource at low damping, at known points}
Carrying out PBT using a resource consisting of $N$ copies of the state in Eq.~(\ref{eq:new_res}) (which we will call the alternate resource) results in the Choi matrix given in Eq.~(\ref{eq:new_choi}). The difference between Choi matrices with the AD channel is (as given in the main text)
\begin{align}
PBT\left[R_{\mathrm{new}}(a)^{\otimes N}\right]-R'(p_0)=\begin{pmatrix}
x-\frac{1}{2} &0 &0 &z-\frac{\sqrt{1-p_0}}{2}\\
0 &\frac{1}{2}-x &0 &0\\
0 &0 &y-\frac{p_0}{2} &0\\
z-\frac{\sqrt{1-p_0}}{2} &0 &0 &\frac{p_0}{2}-y
\end{pmatrix},
\end{align}
with $x$, $y$ and $z$ defined in the main text. We define $a^{known}$ as the value of $a$ such that the first diagonal element of this matrix is the same as the third diagonal element. This is a value of $a$ for which the diamond norm is known analytically and is equal to the trace norm between Choi matrices; we refer to this as a known point. At the point $a^{known}=\frac{1}{2}$ the resource state is simply a maximally entangled state.

Carrying out PBT using a resource consisting of $N$ copies of the state in Eq.~(\ref{eq:choi_res}) (which we will call the Choi resource) results in the Choi matrix given in Eq.~(\ref{eq:choi of choi_res}), and the difference between Choi matrices is (as given in the main text)
\begin{align}
PBT[R(p_1)^{\otimes N}]-R'(p_0)=\begin{pmatrix}
-e_1 &0 &0 &-c\\
0 &e_1 &0 &0\\
0 &0 &e_2 &0\\
-c &0 &0 &-e_2
\end{pmatrix},
\end{align}
with $e_1$, $e_2$ and $c$ defined in the main text. We define $p_1^{known}$ as the value of $p_1$ such that the first diagonal element of this matrix is the same as the third diagonal element, similarly to $a^{known}$. The minimum value of $p_1^{known}$ is 0; at this point the resource state is again a maximally entangled state.

The corresponding $p_0$ value for $a^{known}=\frac{1}{2}$ is $\frac{\xi_N}{2}$. The corresponding $p_0$ value for $p_1^{known}=0$ is also $\frac{\xi_N}{2}$. Consequently, at this point, both resources simulate the AD channel equally well. Differentiating the expression in Eq.~(\ref{eq:choi_diamond}), we find that the gradient of the diamond norm for the Choi resource at $p_1=p_1^{known}$, $D_{\diamond}^1$, is
\begin{align}
\frac{d D_{\diamond}^1}{d p_0}=-\frac{1}{2}\left(\frac{\xi_N}{1-\xi_N}+\frac{2\left(1-\sqrt{1-\xi_N}\right)^2+\frac{(1-p_0)\xi_N^2}{(1-\xi_N)^2}}{\sqrt{4(1-p_0)\left(1-\sqrt{1-\xi_N}\right)^2+\frac{(1-p_0)^2\xi_N^2}{(1-\xi_N)^2}}}\right),
\end{align}
which is finite and negative for all $\xi_N<1$ (a condition which holds for all $N\geq2$). We will now show that the gradient of the diamond norm for the alternate resource at $a=a^{known}$, which we will denote as $D_{\diamond}^2$, diverges as $a^{known}$ tends to $\frac{1}{2}$ from above.

We first find that $D_{\diamond}^2$ takes the form
\begin{align}
D_{\diamond}^2=\left. p_0-2y+\sqrt{(p_0-2y)^2+(\sqrt{1-p_0}-2z)^2}\right|_{a=a^{known}(p_0)},
\end{align}
by using the fact that the eigenvalues of a matrix of the form
\begin{align}
\begin{pmatrix}
x_1 &0 &0 &x_2\\
0 &-x_1 &0 &0\\
0 &0 &x_1 &0\\
x_2 &0 &0 &-x_1
\end{pmatrix}
\end{align}
are $\{\pm x_1,\sqrt{x_1^2+x_2^2}\}$. We then differentiate $D_{\diamond}^2$, getting
\begin{align}
\begin{split}
\frac{d D_{\diamond}^2}{d p_0}&=1-2\frac{dy}{da}\frac{da^{known}}{dp_0}+\frac{(p_0-2y)\left(1-2\frac{dy}{da}\frac{da^{known}}{dp_0}\right)+(\sqrt{1-p_0}-2z)\left(\frac{-1}{2\sqrt{1-p_0}}-2\frac{dz}{da}\frac{da^{known}}{dp_0}\right)}{\sqrt{(p_0-2y)^2+(\sqrt{1-p_0}-2z)^2}}\\
&=\left(1+\frac{(p_0-2y)-\frac{1}{2}+\frac{2z}{2\sqrt{1-p_0}}}{\sqrt{(p_0-2y)^2+(\sqrt{1-p_0}-2z)^2}}\right)-2\frac{da^{known}}{dp_0}\left(\frac{dy}{da}+\frac{(p_0-2y)\frac{dy}{da}+(\sqrt{1-p_0}-2z)\frac{dz}{da}}{\sqrt{(p_0-2y)^2+(\sqrt{1-p_0}-2z)^2}}\right)\label{eq:dif_of_alt_diamond}
\end{split}
\end{align}
where $y$ and $z$ are evaluated at $a=a^{known}(p_0)$. We will show that the term in the right-hand bracket of Eq.~(\ref{eq:dif_of_alt_diamond}) is positive sufficiently close to $a=\frac{1}{2}$. Note that since $x\leq\frac{1}{2}$ and $x-\frac{1}{2}=y-\frac{p_0}{2}$, $p_0-2y\geq0$

Let us find an expression for $\frac{dy}{da}$. Recall that $y$ is given by
\begin{align}
\begin{split}
y=\sum_{s=s_{\mathrm{min}}}^{\frac{N-1}{2}}\sum_{m=-s}^{s}a^{\frac{N-1}{2}+m}(1-a)^{\frac{N+1}{2}-m}\frac{N!(s+m)(s-m+1)\left[\left(\frac{N+1}{2}-s\right)^{-\frac{1}{2}}-\left(\frac{N+3}{2}+s\right)^{-\frac{1}{2}}\right]^2}{2\left(\frac{N-1}{2}-s\right)!\left(\frac{N+1}{2}+s\right)!(2s+1)}\\
+\sum_{m=-\frac{N+1}{2}}^{\frac{N+1}{2}}a^{\frac{N-1}{2}+m}(1-a)^{\frac{N+1}{2}-m}\frac{\left(\frac{N-1}{2}+m\right)\left(\frac{N+1}{2}+m\right)}{2N(N+1)},
\end{split}
\end{align}
and define $\textrm{cont}_1^y(s,m)$ and $\textrm{cont}_2^y(m)$ such that
\begin{align}
y=\sum_{s=s_{\mathrm{min}}}^{\frac{N-1}{2}}\sum_{m=-s}^{s}a^{\frac{N-1}{2}+m}(1-a)^{\frac{N+1}{2}-m}\textrm{cont}_1^y(s,m)+\sum_{m=-\frac{N+1}{2}}^{\frac{N+1}{2}}a^{\frac{N-1}{2}+m}(1-a)^{\frac{N+1}{2}-m}\textrm{cont}_2^y(m),
\end{align}
noting that $\textrm{cont}_1^y(s,m)$ and $\textrm{cont}_2^y(m)$ have no $a$-dependence. Hence, applying the product rule of differentiation,
\begin{align}
\begin{split}
\frac{dy}{da}=\frac{N(1-2a)}{2a(1-a)}y+\sum_{s=s_{\mathrm{min}}}^{\frac{N-1}{2}}\sum_{m=-s}^{s}a^{\frac{N-1}{2}+m}(1-a)^{\frac{N+1}{2}-m}\frac{2m-1}{2a(1-a)}\textrm{cont}_1^y(s,m)+\\
\sum_{m=-\frac{N+1}{2}}^{\frac{N+1}{2}}a^{\frac{N-1}{2}+m}(1-a)^{\frac{N+1}{2}-m}\frac{2m-1}{2a(1-a)}\textrm{cont}_2^y(m).
\end{split}\label{eq:y_diff}
\end{align}
Note that if $m$ goes to $1-m$, $\textrm{cont}_1^y(s,m)$ is unchanged (i.e. $\textrm{cont}_1^y(s,m)=\textrm{cont}_1^y(s,1-m)$) and $2m-1$ goes to $-(2m-1)$. Note too that $\textrm{cont}_1^y(s,-s)=0$ and that $m=\frac{1}{2}$ sets $2m-1$ to 0, meaning that we can write
\begin{align}
\begin{split}
\frac{dy}{da}=\frac{N(1-2a)}{2a(1-a)}y+\sum_{s=s_{\mathrm{min}}}^{\frac{N-1}{2}}\sum_{m=\{1,\frac{3}{2}\}}^{s}\left(a^{\frac{N-1}{2}+m}(1-a)^{\frac{N+1}{2}-m}-a^{\frac{N+1}{2}-m}(1-a)^{\frac{N-1}{2}+m}\right)\frac{2m-1}{2a(1-a)}\textrm{cont}_1^y(s,m)+\\
\sum_{m=\{1,\frac{3}{2}\}}^{\frac{N+1}{2}}\left(a^{\frac{N-1}{2}+m}(1-a)^{\frac{N+1}{2}-m}\textrm{cont}_2^y(m)-a^{\frac{N+1}{2}-m}(1-a)^{\frac{N-1}{2}+m}\textrm{cont}_2^y(1-m)\right)\frac{2m-1}{2a(1-a)},
\end{split}
\end{align}
where the minimum value of $m$ is 1 for odd $N$ and $\frac{3}{2}$ for even $N$. We now note that, for $a\geq\frac{1}{2}$,
\begin{align}
a^{\frac{N-1}{2}+m}(1-a)^{\frac{N+1}{2}-m} \geq a^{\frac{N+1}{2}-m}(1-a)^{\frac{N-1}{2}+m},
\end{align}
with equality only at $a=\frac{1}{2}$, meaning that sufficiently close to $a=\frac{1}{2}$, the second sum in Eq.~(\ref{eq:y_diff}) dominates. Note too that $\textrm{cont}_2^y(m)>\textrm{cont}_2^y(1-m)$ (with a finite difference between $\textrm{cont}_2^y(m)$ and $\textrm{cont}_2^y(1-m)$ that does not depend on $a$), and hence $\frac{dy}{da}>0$ for $a$ sufficiently close to $\frac{1}{2}$.

Let us now find an expression for $\frac{dz}{da}$. Recall that $z$ is given by
\begin{align}
\begin{split}
z=\sum_{s=s_{\mathrm{min}}}^{\frac{N-1}{2}}\sum_{m=-s}^{s}&\frac{a^{\frac{N}{2}+m}(1-a)^{\frac{N}{2}-m}N!}{2\left(\frac{N-1}{2}-s\right)!\left(\frac{N+1}{2}+s\right)!(2s+1)}\left[\left(\frac{N+1}{2}-s\right)^{-1}(s^2-m^2)\right.\\
&\left.+2\left(\frac{N+1}{2}-s\right)^{-\frac{1}{2}}\left(\frac{N+3}{2}+s\right)^{-\frac{1}{2}}(s^2+m^2+s)+\left(\frac{N+3}{2}+s\right)^{-1}((s+1)^2-m^2)\right]\\
&-\sum_{m=-\frac{N+1}{2}}^{\frac{N+1}{2}}a^{\frac{N}{2}+m}(1-a)^{\frac{N}{2}-m}\frac{\left(\frac{N+1}{2}+m\right)\left(\frac{N+1}{2}-m\right)}{2N(N+1)},
\end{split}
\end{align}
and define $\textrm{cont}_1^z(s,m)$ and $\textrm{cont}_2^z(m)$ such that
\begin{align}
z=\sum_{s=s_{\mathrm{min}}}^{\frac{N-1}{2}}\sum_{m=-s}^{s}a^{\frac{N}{2}+m}(1-a)^{\frac{N}{2}-m}\textrm{cont}_1^z(s,m)+\sum_{m=-\frac{N+1}{2}}^{\frac{N+1}{2}}a^{\frac{N}{2}+m}(1-a)^{\frac{N}{2}-m}\textrm{cont}_2^z(m).
\end{align}
Differentiating, we get
\begin{align}
\begin{split}
\frac{dz}{da}=\frac{N(1-2a)}{2a(1-a)}z+\sum_{s=s_{\mathrm{min}}}^{\frac{N-1}{2}}\sum_{m=-s}^{s}a^{\frac{N}{2}+m}(1-a)^{\frac{N}{2}-m}\frac{m}{a(1-a)}\textrm{cont}_1^z(s,m)+\\
\sum_{m=-\frac{N+1}{2}}^{\frac{N+1}{2}}a^{\frac{N}{2}+m}(1-a)^{\frac{N}{2}-m}\frac{m}{a(1-a)}\textrm{cont}_2^z(m).
\end{split}
\end{align}
Note that $\textrm{cont}_1^z(s,m)=\textrm{cont}_1^z(s,-m)$ and $\textrm{cont}_2^z(s,m)=\textrm{cont}_2^z(s,-m)$. Hence, we can write
\begin{align}
\begin{split}
\frac{dz}{da}=\frac{N(1-2a)}{2a(1-a)}z+\sum_{s=s_{\mathrm{min}}}^{\frac{N-1}{2}}\sum_{m=\{1,\frac{3}{2}\}}^{s}\left(a^{\frac{N}{2}+m}(1-a)^{\frac{N}{2}-m}-a^{\frac{N}{2}-m}(1-a)^{\frac{N}{2}+m}\right)\frac{m}{a(1-a)}\textrm{cont}_1^z(s,m)+\\
\sum_{m=\{1,\frac{3}{2}\}}^{\frac{N+1}{2}}\left(a^{\frac{N}{2}+m}(1-a)^{\frac{N}{2}-m}-a^{\frac{N}{2}-m}(1-a)^{\frac{N}{2}+m}\right)\frac{m}{a(1-a)}\textrm{cont}_2^z(s,m).
\end{split}
\end{align}
Note that this approaches 0 as $a$ approaches $\frac{1}{2}$, hence there exists some finite, positive $\epsilon$ such that for all $\frac{1}{2}\leq a \leq \frac{1}{2}+\epsilon$, we have
\begin{align}
\frac{dy}{da}+\frac{(p_0-2y)\frac{dy}{da}+(\sqrt{1-p_0}-2z)\frac{dz}{da}}{\sqrt{(p_0-2y)^2+(\sqrt{1-p_0}-2z)^2}}>0.
\end{align}

It now suffices to show that $\frac{da^{known}}{dp_0}$ diverges as $a$ tends to $\frac{1}{2}$ from above. We write
\begin{align}
\frac{da^{known}}{dp_0}=\left(\frac{dp_0}{da^{known}}\right)^{-1}=\frac{d}{da}\left(1-2(x-y)\right)=-2\frac{d}{da}(x-y).
\end{align}
Using the symmetry of the PBT protocol, we can see that $x[a]=\frac{1}{2}-y[1-a]$. We can therefore write
\begin{align}
\frac{dp_0}{da^{known}}=2\frac{d}{da}\left(y[a]+y[1-a]\right).\label{eq:p0_diff}
\end{align}
The differential $\frac{dy[a]}{da}$ is given in Eq.~(\ref{eq:y_diff}), and we can similarly write
\begin{align}
\begin{split}
\frac{dy[1-a]}{da}=&\frac{N(1-2a)}{2a(1-a)}y[1-a]+\\
&\sum_{s=s_{\mathrm{min}}}^{\frac{N-1}{2}}\sum_{m=\{1,\frac{3}{2}\}}^{s}\left(a^{\frac{N-1}{2}+m}(1-a)^{\frac{N+1}{2}-m}-a^{\frac{N+1}{2}-m}(1-a)^{\frac{N-1}{2}+m}\right)\frac{2m-1}{2a(1-a)}\textrm{cont}_1^y(s,m)+\\
&\sum_{m=\{1,\frac{3}{2}\}}^{\frac{N+1}{2}}\left(a^{\frac{N-1}{2}+m}(1-a)^{\frac{N+1}{2}-m}\textrm{cont}_2^y(1-m)-a^{\frac{N+1}{2}-m}(1-a)^{\frac{N-1}{2}+m}\textrm{cont}_2^y(m)\right)\frac{2m-1}{2a(1-a)}.
\end{split}
\end{align}
The expression $y[a]+y[1-a]$ is symmetric around $a=\frac{1}{2}$ and both $\frac{dy[a]}{da}$ and $\frac{dy[1-a]}{da}$ are finite at this point, so $a=\frac{1}{2}$ is either a maximum or a minimum of this expression.

Suppose that it is a minimum. Numerically, we find a clear trend indicating that this is the case for all $N$, with the second differential tending towards 1 from below (from a value of 0 at $N=2$) as $N$ increases. Then, $\frac{da^{known}}{dp_0}$ diverges to positive infinity as $a$ approaches $\frac{1}{2}$ from above. Consequently, $\frac{d D_{\diamond}^2}{d p_0}$ diverges to negative infinity. Hence, there exists some finite positive $\epsilon$ such that the gradient of the diamond norm for the Choi resource, assessed at $p_0=\frac{\xi_N}{2}+\delta$ is less negative than the gradient of the diamond norm for the alternate resource, assessed at the same point, for all positive $\delta<\epsilon$. Consequently, the diamond norm for the Choi resource at the known point is less than the diamond norm for the alternate resource for all $\frac{\xi_N}{2}<p_0\leq\epsilon$.

Suppose instead that it is a maximum. Then, $\frac{da^{known}}{dp_0}$ diverges to negative infinity as $a$ approaches $\frac{1}{2}$ from above, and $\frac{d D_{\diamond}^2}{d p_0}$ diverges to positive infinity. However, in this case, increasing $a$ by a small amount from $\frac{1}{2}$ decreases $p_0$, since $\frac{dp_0}{da^{known}}$ is negative. Consequently, there exists some finite positive $\epsilon$ such that $D_{\diamond}^2$ assessed at $p_0=\frac{\xi_N}{2}-\delta$ is lower than  $D_{\diamond}^1$ assessed at $p_0=\frac{\xi_N}{2}+\delta$ for all positive $\delta<\epsilon$. In this case, an AD channel applied to the output of the PBT channel, with the damping probability $p'$ chosen such that total channel simulates an AD channel with $p_0=\frac{\xi_N}{2}+\delta$ would result in $D_{\diamond}^2<D_{\diamond}^1$. This is equivalent to using the tensor-product resource composed of $N$ copies of
\begin{align}
R_{\mathrm{new}}'(a)=\begin{pmatrix}
a(1-p') &0 &0 &0\\
0 &a &-\sqrt{a(1-a)(1-p')} &0\\
0 &-\sqrt{a(1-a)(1-p')} &(1-a)(1-p') &0\\
0 &0 &0 &0
\end{pmatrix},
\end{align}
which is still distinct from any state of the form in Eq.~(\ref{eq:choi_res}).

Hence, in either case and for any $N$, there exists some tensor-product resource that simulates the AD channel better than the Choi resource, at either of its known points, for some range of $p_0$ values.

\end{document}